%% file: main.tex
\documentclass{article}

\usepackage{microtype}
\usepackage{graphicx}
\usepackage{subfigure}
\usepackage{cancel}
\usepackage{booktabs} 
\usepackage{eso-pic}

\usepackage{subcaption}
\usepackage{booktabs} 
\usepackage{color}
\usepackage{tcolorbox}
\usepackage[table]{xcolor}

\usepackage{hyperref}

\usepackage[accepted]{mlsys2025}

\usepackage{graphicx} 
\usepackage{amsmath}
\usepackage{amsfonts}
\usepackage{booktabs}
\usepackage{makecell}
\usepackage{multirow}
\usepackage{amssymb}
\usepackage{breakurl}

\usepackage[font=footnotesize]{caption}

\usepackage{adjustbox}
\usepackage{listings}
\usepackage[hang,flushmargin]{footmisc}

\usepackage{tikz}

\usepackage{enumitem}

\usepackage{graphicx}
\usepackage{hyperref}
\usepackage{eso-pic}
\usepackage{xparse}
\usepackage{xspace}
\usepackage{titlesec}

\newcommand{\SysName}{Spira\xspace}
\newcommand{\SpConv}{SpC\xspace}

\newcommand{\SpiraEndToEndAvgAll}{1.68$\times$\xspace}

\newcommand{\SpiraEndToEndMaxAll}{3.04$\times$\xspace}

\newcommand{\SpiraEndToEndAvgThirty}{1.70$\times$\xspace} 

\newcommand{\SpiraEndToEndAvgAHundred}{2.05$\times$\xspace} 

\newcommand{\SpiraLayerAvgThirty}{2.11$\times$\xspace} 

\newcommand{\SpiraLayerMaxThirty}{3.44$\times$\xspace} 

\newcommand\cg[1]{\noindent{\color{black} {#1}}} 
\newcommand\da[1]{\noindent{\color{black} {#1}}} %

\newcommand\camera[1]{\noindent{\color{black}{#1}}}

\mlsystitlerunning{\SysName: Exploiting Voxel Data Structural Properties  for Efficient Sparse Convolution in Point Cloud Networks}

\begin{document}
\titlespacing*{\section}{0pt}{4pt}{2pt}
\titlespacing*{\subsection}{0pt}{2pt}{1pt}
\titlespacing*{\subsubsection}{0pt}{2pt}{2pt}

\twocolumn[
\mlsystitle{\SysName: Exploiting Voxel Data Structural Properties \\ for Efficient Sparse Convolution in Point Cloud Networks}

\AddToShipoutPictureFG*{%
  \AtPageUpperLeft{%
    \hspace*{\dimexpr\paperwidth-8.5cm\relax}%
    \raisebox{-2.4cm}[0pt][0pt]{%
      \makebox[0pt][l]{%
        \href{https://www.acm.org/publications/policies/artifact-review-and-badging-current}{%
          \includegraphics[width=54pt]{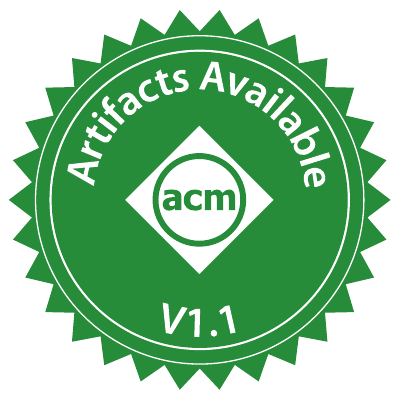}%
        }\hspace{8pt}%
        \href{https://www.acm.org/publications/policies/artifact-review-and-badging-current}{%
          \includegraphics[width=54pt]{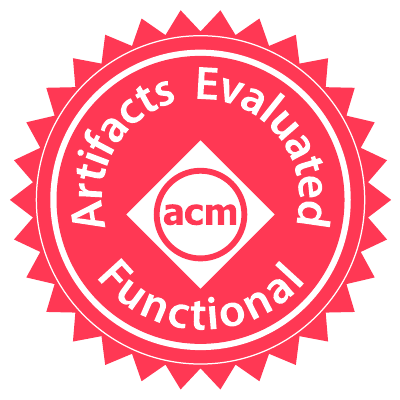}%
        }\hspace{8pt}%
        \href{https://www.acm.org/publications/policies/artifact-review-and-badging-current}{%
          \includegraphics[width=54pt]{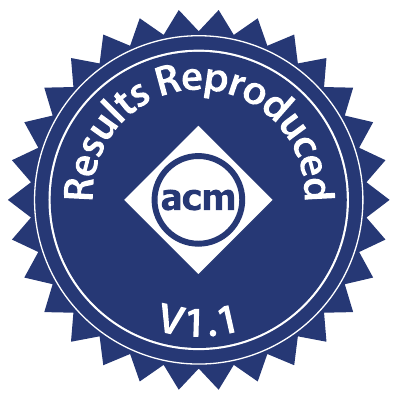}%
        }%
      }%
    }%
  }%
}




\begin{mlsysauthorlist}
\mlsysauthor{Dionysios Adamopoulos}{MPI,NTUA}
\mlsysauthor{Anastasia Poulopoulou}{NTUA}
\mlsysauthor{Georgios Goumas}{NTUA}
\mlsysauthor{Christina Giannoula}{MPI}
\end{mlsysauthorlist}

\mlsysaffiliation{MPI}{SPIN Research Group, Max Planck Institute for Software Systems (MPI-SWS)}
\mlsysaffiliation{NTUA}{CSLab Research Group, National Technical University of Athens (NTUA)}

\mlsyscorrespondingauthor{Christina Giannoula}{cgiannoula@mpi-sws.org}


\mlsyskeywords{Machine Learning, MLSys}

\vspace{14pt}
\input{content/abstract}
]



\printAffiliationsAndNotice{}  

\input{content/introduction}

\input{content/background}

\input{content/spconv_engines}

\input{content/voxel_properties}

\input{content/mechanism}

\input{content/evaluation}

\input{content/related_work}
\input{content/conclusion}

\input{content/acknowledgements}

\nocite{langley00}

\bibliography{references}
\bibliographystyle{mlsys2025}


\newpage
\appendix
\input{content/appendix}

\end{document}

%% file: content/abstract.tex
\begin{abstract}

Sparse Convolution (\SpConv) \cg{powers} 3D point cloud networks widely used in autonomous driving and \camera{augmented/virtual reality}. \SpConv builds a kernel map that stores mappings between input voxel coordinates, output coordinates, and weight offsets, then uses this map to compute feature vectors for output coordinates. Our work identifies three key properties of voxel coordinates: they are integer-valued, bounded within a limited spatial range, and geometrically continuous, i.e., neighboring voxels on the same object surface are highly likely to exist at small spatial offsets from each other. Prior \SpConv engines do not fully exploit these properties and suffer from high pre-processing and post-processing overheads during kernel map construction.
To this end, we design \SysName, the first voxel-property-aware \SpConv engine for GPUs. \SysName proposes (i) a high-performance one-shot search algorithm that builds the kernel map with \emph{no} pre-processing and high \camera{data} locality, (ii) an effective packed-native processing scheme that accesses packed voxel coordinates at low cost, (iii) a flexible dual-dataflow execution mechanism that efficiently computes output feature vectors by adapting to layer characteristics, and (iv) a network-wide parallelization strategy that builds kernel maps for all \SpConv layers concurrently at network start.
Our evaluation shows that \SysName significantly outperforms prior \camera{state-of-the-art} \SpConv engines by \camera{\SpiraEndToEndAvgAll} on average and up to \camera{\SpiraEndToEndMaxAll} for end-to-end inference, and by \camera{\SpiraLayerAvgThirty} on average and up to \camera{\SpiraLayerMaxThirty} for layer-wise execution across diverse layer configurations. \camera{The source code of \SysName is freely
available at \href{https://github.com/SPIN-Research-Group/Spira}{https://github.com/SPIN-Research-Group/Spira}}.
\end{abstract}

%% file: content/introduction.tex
\section{Introduction}

Point cloud data has become increasingly used in various important applications, including autonomous driving~\cite{zermas2017fast}, robotics~\cite{kim2018slam}, augmented/virtual reality~\cite{wirth2019pointatme}, and drones~\cite{zheng2020point}. Light detection and ranging (LiDAR) sensors on autonomous vehicles, drones, and mobile devices generate point cloud data that \camera{is} processed by point cloud networks. Voxel-based point cloud networks achieve state-of-the-art accuracy~\cite{hong2023exploiting} in vision tasks such as object detection~\cite{yin2021center} and segmentation~\cite{zhu2021cylindrical} by processing voxelized point clouds (referred to as \textbf{voxel data})—raw point cloud data have been quantized into a discrete 3D grid of small cubes (voxels).

Unlike 2D images, 3D point cloud \camera{data is} extremely sparse, typically occupying \camera{less than 1\% of its bounding volume ~\cite{hong2023exploiting}}.
Sparse Convolution (\textbf{\SpConv}) is therefore the dominant computational kernel in voxel-based point cloud networks, consisting of two steps. 1) The \textbf{voxel indexing} step generates the output voxel coordinates,  and the mappings between the input coordinates,  output coordinates and the weight offsets. Voxel indexing finds and stores these mappings in a table called \textbf{kernel map}, performing lookup operations (\textbf{searches}) on a query data structure containing the voxel coordinates.
To accomplish this, \SpConv engines include a \textbf{pre-processing} phase that organizes coordinates in a query data structure, and may  include a \textbf{post-processing} phase that rearranges and filters the kernel map for the subsequent feature computation step. 2) The \textbf{feature computation} step produces the feature vectors (actual convolution output) for the output coordinates based on the kernel map mappings. This step is parallelized across thread blocks using one of two  dataflow approaches: \textbf{output-stationary} or \textbf{weight-stationary}, explained in §\ref{sec:background-execution-steps}. Prior work~\cite{Tang2023TorchSparse++} shows that both dataflows are necessary, since each performs best under different layer characteristics. 

In this work, we extensively characterize voxel data processed in \SpConv and analyze prior state-of-the-art \SpConv engines. 
We identify two key limitations (§\ref{sec:spconv-engines-limitations}): prior works~\cite{tang2022torchsparse,Tang2023TorchSparse++,Yang2024Minuet,spconv2022,hong2023exploiting,Choy2019Minkownski} incur non-negligible performance overheads in the pre-processing and post-processing phases of voxel indexing, and lack efficient support for both output- and weight-stationary dataflows required for feature computation. 
We also \camera{discover} three key properties of voxel data. 
First, voxel coordinates are integer-valued, as \cg{each voxel point is a triplet of values representing a \emph{discrete} voxel (cube) in a 3D grid.}  
Second, they are spatially bounded within a limited value range, since they represent a \emph{finite} 3D grid, the exact size of which depends on the captured environment of each application (§\ref{sec:voxel-properties}). 
Third, they are geometrically continuous, as they represent continuous object surfaces: neighboring voxel coordinates on the same object are highly likely to exist at small offset displacements from each other.
Prior \SpConv engines do \emph{not} exploit these properties. We demonstrate that by leveraging these properties we can significantly reduce computation and data access costs in \SpConv, leading to substantial performance improvements.

To this end, we design \SysName, the first \SpConv engine that comprehensively exploits \camera{structural properties of the} voxel data to accelerate performance.
\SysName relies on four key ideas.
First, we design a \emph{one-shot} search algorithm that \emph{completely} eliminates the pre-processing phase in voxel indexing, and performs fast localized searches by exploiting the integer-valued property, thereby reducing expensive irregular lookup operations.
Second, \cg{we pack each voxel point triplet into a single integer value by exploiting the spatially bounded property of voxel data,} and design packed-native kernels that \emph{directly} perform voxel indexing on packed data, eliminating unpacking/repacking overheads and reducing compute and data access costs. 
Third, we propose an adaptive hybrid dataflow scheme for feature computation that selects output-stationary, weight-stationary dataflow or hybrid combinations of both dataflows based on kernel map density patterns identified using the geometric continuity property of voxel data. 
This approach improves feature computation performance by adapting execution dataflow to both layer characteristics and kernel map densities.
Fourth, we identify that voxel indexing steps across \SpConv layers have \emph{no} dependencies between them or with other operators  throughout the network, and we execute all voxel indexing steps concurrently at the network start across multiple GPU SMs, thus improving resource utilization and execution parallelism.


We extensively evaluate \SysName using a wide variety of 3D point cloud networks, real-world datasets from indoor and outdoor scenes, and \camera{six} GPU \camera{systems}, \camera{spanning from high-end to edge  GPU architectures. We demonstrate}  that \SysName significantly outperforms prior \SpConv engines. In end-to-end point cloud inference, \SysName improves performance by \camera{\SpiraEndToEndAvgAll on average (up to \SpiraEndToEndMaxAll)}. 
At the \SpConv layer, \SysName achieves \camera{\SpiraLayerAvgThirty} speedup on average and up to \camera{\SpiraLayerMaxThirty} speedup over prior work across various layer configurations. 

We make the following contributions in this work: 
\begin{itemize}[topsep=0pt,leftmargin=8pt,nosep,partopsep=0pt,before=\vspace{-\parskip}]
    \item We identify important structural properties of voxel data and propose \SysName, the first voxel-property-aware \SpConv engine for GPU \camera{systems}.
    \item We propose a one-shot search algorithm that performs localized searches and eliminates pre-processing, and design a packed-native voxel indexing \camera{scheme} that exploits voxel properties to reduce compute and data access costs. We propose network-wide voxel indexing \camera{execution} that executes all layers' voxel indexing steps concurrently, and design an adaptive hybrid dataflow scheme that adapts feature computation dataflow to layer characteristics and kernel map densities. 
    \item We evaluate \SysName across diverse real-world datasets, point cloud networks and \camera{six} GPU \camera{systems (ranging from high-end to edge GPUs)}, demonstrating significant performance improvements over state-of-the-art \SpConv engines. 
    \item \camera{We open-source \SysName in our GitHub repository: \href{https://github.com/SPIN-Research-Group/Spira}{https://github.com/SPIN-Research-Group/Spira}}.
\end{itemize}

%% file: content/background.tex
\section{Background}
\subsection{Sparse Convolution (\SpConv) Definition}\label{sec:background-definition}

A \textit{point cloud} is an unordered set of 3D points,  representing locations on the surface of an object or within a scene. Unlike images, point clouds are irregularly distributed and spatially sparse, making standard convolutional neural networks inefficient~\cite{Wu2019PointConv}. To enable convolutional processing, point clouds are often voxelized by discretizing the 3D space into a regular grid of small cubes (voxels). A voxelized point cloud (henceforth referred as \textbf{voxel data}) can be represented as a set of tuples $\{(\mathbf{v}_i, \mathbf{f}_i)\}$, where $\mathbf{v}_i \in \mathbb{Z}^3$ denotes the quantized coordinate of the $i$-th voxel in the $3D$-grid space, and $\mathbf{f}_i \in \mathbb{R}^C$ is its corresponding feature vector. The quantization from continuous to discrete space is performed as $\mathbf{v}_i = \lfloor \mathbf{p}_i^{(\text{raw})} / \mathbf{g} \rfloor$, where $\mathbf{p}_i^{(\text{raw})} \in \mathbb{R}^3$ are the original continuous coordinates of the $i$-th point and $\mathbf{g} \in \mathbb{R}^3$  the grid size vector that determines the spatial resolution, both typically measured in meters. 

Sparse Convolution (\SpConv) is the dominant operation in voxel-based point cloud networks. It takes as input voxel data \cg{$p$} with coordinates $\mathbf{V}_p$  and its output \cg{$q$} is also a voxel data with coordinates $\mathbf{V}_q$, controlled by a stride parameter ~\cite{Choy2019Minkownski}. 
Given  input stride $s_p$ and layer stride $s_l$, the output stride is $s_q = s_p \times s_l$. When $s_l = 1$, the output coordinates coincide with the input, $\mathbf{V}_q = \mathbf{V}_p$ (\textbf{submanifold} convolution), which comprises the majority of layers in point cloud networks.  
When $s_l > 1$, \textit{downsampling} is applied for the output coordinates as $\mathbf{V}_q = \lfloor \frac{\mathbf{V}_p}{s_q} \rfloor \times s_q (1)\label{eq:1}$, keeping only the unique values. Downsampling typically uses $s_l = 2$, so after $m$ downsamplings, \cg{the output stride becomes $s_q = 2^m$~\cite{Lin2021PointAcc}.}

For the input features $F_p \in \mathbb{R}^{|\mathbf{V}_p| \times \mathbf{C_{in}}}$ and the output features $F_q \in \mathbb{R}^{|\mathbf{V}_q| \times \mathbf{C_{out}}}$, $\mathbf{C_{in}}$ and $\mathbf{C_{out}}$ are the input and output channel sizes, respectively. The output feature vector $\mathbf{f}^q_i$ of the $i$-th output coordinate $\mathbf{q}_i$ is computed over all weight offsets $\boldsymbol{\delta}_k$ and input coordinates $\mathbf{p}_j$ that satisfy the condition $\mathbf{p}_j = \mathbf{q}_i + \boldsymbol{\delta}_k$, as follows:
\label{eq:2}\[
\mathbf{f}^q_i = \sum_{\boldsymbol{\delta}_k \in \Delta(K,s_p)} \sum_{\mathbf{p}_j \in \mathbf{V}_p} \mathbf{1}_{\mathbf{p}_j = \mathbf{q}_i + \boldsymbol{\delta}_k} \, \mathbf{f}^p_j \, \mathbf{W}_{\boldsymbol{\delta}_k}, \quad (\mathbf{q}_i \in \mathbf{V}_q) (2)
\]

\vspace{-5pt}

where $\Delta(K,s_p)$ is the set of $K^3$ weight offsets of kernel size $K$ and input stride $s_p$ (e.g., $\Delta(5,2) = \{ -4, -2, 0, 2, 4 \}^3$), $\boldsymbol{\delta}_k$ is the $k$-th weight offset, $\mathbf{f}^p_j$ is the feature vector of the input coordinate $\mathbf{p}_j$, $\mathbf{W}_{\boldsymbol{\delta}_k} \in \mathbb{R}^{\mathbf{C_{in}} \times \mathbf{C_{out}}}$ is the weight corresponding to offset $\boldsymbol{\delta}_k$, and $\mathbf{1}_{\mathbf{p}_j = \mathbf{q}_i + \boldsymbol{\delta}_k}$ is the indicator function for the condition $\mathbf{p}_j = \mathbf{q}_i + \boldsymbol{\delta}_k$.

\subsection{The Execution Steps of \SpConv}\label{sec:background-execution-steps}

The execution of a \SpConv layer can be split into two steps: \newline
\noindent\textbf{1) Voxel Indexing Step.} It consists of the \textbf{downsampling} and \textbf{mapping} parts.
Downsampling is performed in \SpConv layers with $s_l$$>$1, and generates output voxel coordinates according to \camera{\hyperref[eq:1]{Equation (1)}} through element-wise rounding and duplicate removal. State-of-the-art downsampling schemes sort coordinates 
to identify and remove duplicates.
\camera{The mapping part} builds a $|\mathbf{V}_q| \times K^3$ matrix $M$, called \emph{\textbf{kernel map}}, which stores the mappings between the input coordinates, the output coordinates and the weight offsets. 
These mappings guide the feature vector computation for output coordinates according to \camera{\hyperref[eq:2]{Equation (2)}}, as described in the  feature computation step.
\SpConv engines build the kernel map in three phases: (i) organize coordinates in a query data structure (\textbf{pre-processing} phase), (ii) perform lookup operations (\textbf{search} phase) of the \emph{queries}  $\mathbf{q_i} + \boldsymbol{\delta}_\mathbf{k}$ for each output coordinate $\mathbf{q_i}$ and each weight offset $\boldsymbol{\delta}_\mathbf{k}$: if the query $\mathbf{q_i} + \boldsymbol{\delta}_\mathbf{k}$ matches an input coordinate $\mathbf{p_j}$ in the query data structure, then $M[i,k]= j$, otherwise $M[i,k]= -1$ (invalid entry), (iii) filter  invalid entries (-1) and rearrange the kernel map (\textbf{post-processing} phase), if needed. Thus, \camera{the} mapping \camera{part} has pre-processing kernels that build the query data structure, search operations and potentially post-processing kernels that rearrange and filter the kernel map for the feature computation \camera{step}.

\begin{figure}[t]  
    \centering
    \includegraphics[width=\linewidth]{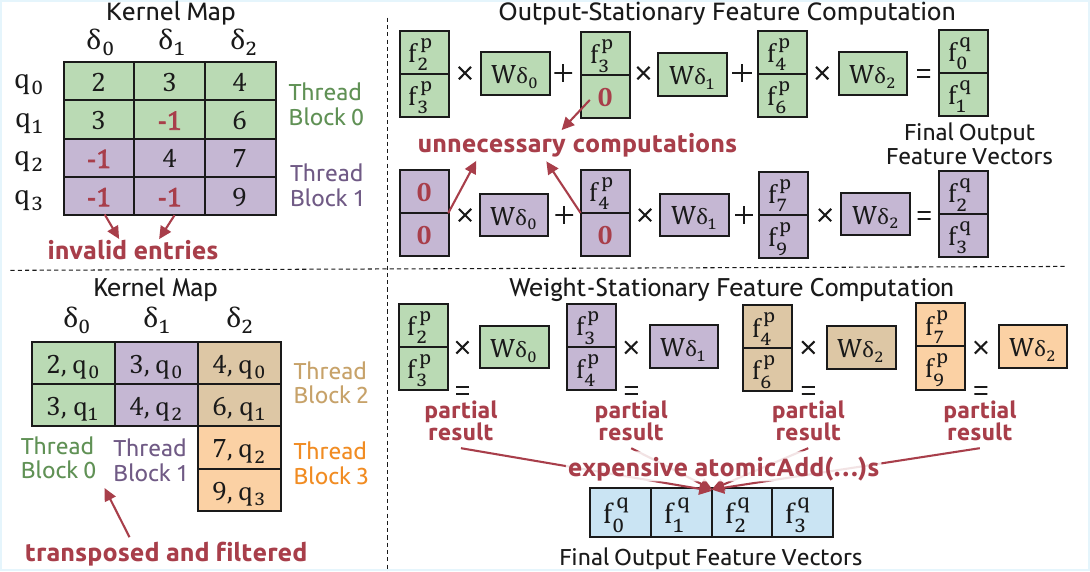}
    \vspace{-20pt}
    \caption{The two dataflows of the feature computation step.}
    \label{fig:background-dataflows}
\end{figure}

\noindent\textbf{2) Feature Computation Step.} It computes the  feature vectors of  output coordinates according to \camera{\hyperref[eq:2]{Equation (2)}}. \camera{There are two types of dataflow}, shown in \autoref{fig:background-dataflows}. \textbf{I) The output-stationary} dataflow (\autoref{fig:background-dataflows} top) distributes the output coordinates $\mathbf{q_i}$ across thread blocks: each thread block computes feature vectors associated with all weight offsets $\boldsymbol{\delta}_\mathbf{k}$ for a chunk of the output coordinates. 
Although thread blocks \emph{directly} produce the final output feature vectors $\mathbf{f}^q_i$, they perform \textbf{unnecessary} zero-valued multiplications, since the kernel map is \emph{not} filtered to remove invalid entries.  
\textbf{II) The weight-stationary} dataflow (\autoref{fig:background-dataflows} bottom) requires the kernel map to be \textbf{transposed} relative to output-stationary format to enable \textbf{coalesced} memory writes among threads within a thread block, and \textbf{filtered} from invalid entries. It distributes the weight offsets across thread blocks: each thread block is assigned a weight offset $\boldsymbol{\delta}_\mathbf{k}$ and computes partial sums for the output feature vectors $\mathbf{f}^q_i$ over a chunk of  valid input–output coordinate pairs. Although this dataflow eliminates unnecessary computations, thread blocks produce \emph{partial results} for \emph{same} output feature vectors, which are merged using \textbf{expensive atomic instructions} (e.g., \texttt{atomicAdd}) to obtain the final output vectors. 
Prior work~\cite{Tang2023TorchSparse++} shows that \textbf{both} dataflows are needed to achieve optimal performance across varying layer configurations and kernel map densities.

%% file: content/spconv_engines.tex
\section{Existing \SpConv Engines}\label{sec:spconv-engines-limitations}

A few prior works improve \SpConv performance on GPUs. MinkowskiEngine~\cite{Choy2019Minkownski} is the first open-source library providing a generalized sparse convolution for point clouds. SpConv2~\cite{spconv2022} library introduced the output-stationary dataflow.
TorchSparse~\cite{tang2022torchsparse} optimized the weight-stationary dataflow, while PCEngine ~\cite{hong2023exploiting} further improved it by dynamically adapting between two weight-stationary variants.
TorchSparse++~\cite{Tang2023TorchSparse++} supports \emph{both} dataflows, and its latest public source code~\cite{TorchSparse++Git} integrates optimizations from both SpConv2 and PCEngine, achieving significant performance improvements over these prior works.
Minuet~\cite{Yang2024Minuet} proposes a binary-search-based algorithm for kernel map construction, but \emph{only} supports  weight-stationary dataflow.
While Minuet outperforms MinkowskiEngine and TorchSparse, it does not compare with TorchSparse++.  TorchSparse++ (latest public code) and Minuet are the two best-performing engines; thus, we focus our evaluation \camera{against} these state-of-the-art baselines.

\begin{figure}[t]  
    \centering 
    \includegraphics[width=\linewidth]{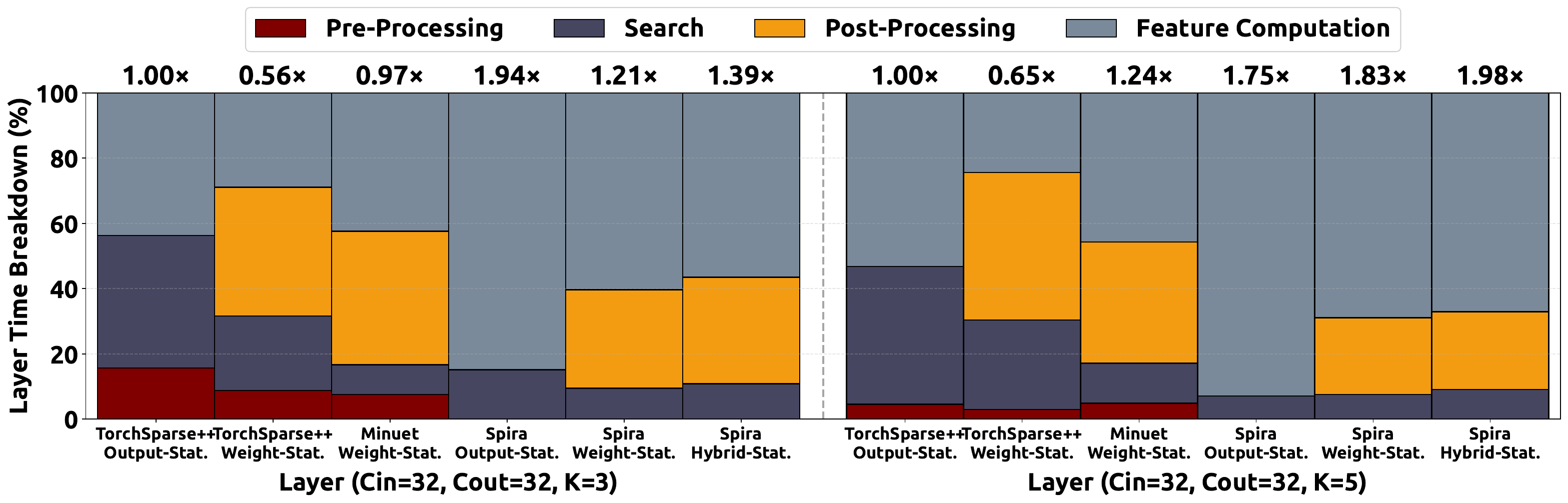}
    \vspace{-22pt}
    \caption{Layer time breakdown using various \SpConv engines. Numbers on bars are speedup over TorchSparse++ output-stationary.}
    \label{fig:motivation-layer-breakdown}
\end{figure}

\autoref{fig:motivation-layer-breakdown} shows the performance breakdown in two submanifold layers of TorchSparse++  output- and weight-stationary dataflows, Minuet 
and \SysName with both dataflows and its own hybrid dataflow. Numbers above each bar show the speedup of each engine over TorchSparse++ output-stationary. We make two observations. First, although Minuet achieves 4.31$\times$ faster search time than TorchSparse++, it incurs noticeable pre-processing overhead, e.g., in the first layer (left), pre-processing time is \emph{nearly equal} to search time. Instead, \SysName \emph{completely} eliminates pre-processing, and achieves 7.83$\times$ and 1.82$\times$ faster search time than TorchSparse++ and Minuet, respectively.
Second,  TorchSparse++ and Minuet have limited dataflow support. Minuet supports only weight-stationary, while TorchSparse++ incurs significant post-processing costs in weight-stationary by transposing the \emph{entire} kernel map to exploit coalesced memory writes. 
\SysName efficiently supports both dataflows: e.g., \SysName output-stationary performs best in the first layer (left), while \SysName weight-stationary significantly outperforms prior  engines in the second layer (right), having 5.42$\times$ lower post-processing overhead than TorchSparse++. 
Additionally, \SysName enables hybrid dual-dataflow execution that exploits voxel data properties to further accelerate feature computation. In the second layer (right), \SysName with hybrid dataflow is  1.98$\times$ and 1.60$\times$ faster than TorchSparse++ and Minuet, respectively.

%% file: content/voxel_properties.tex
\section{Identified Voxel Data Properties}\label{sec:voxel-properties}

We comprehensively analyze the voxel coordinate data in \SpConv layers, and  we identify three key structural properties. 

\noindent\textbf{(1) Integer Property:}  \emph{Voxel coordinates are integer-valued.} 
Raw point cloud coordinates are quantized to a discrete 3D grid with grid size vector $g$ (§\ref{sec:background-definition}), resulting in integer coordinates ($v_x,v_y,v_z$) along each axis ($x, y, z$). Moreover, \camera{\hyperref[eq:1]{Equation (1)}} explains that when voxel coordinates are  downsampled with stride $s$$>$1, they are rounded to \emph{integer multiples of $s$}.

\noindent\textbf{(2) Bounded Property:} \emph{Voxel coordinates are bounded within a limited value range.} Point cloud data, indoor and outdoor scenes, is spatially constrained by the capture environment and sensor capabilities, having a range of $(R_x, R_y, R_z)$ (e.g., expanding through $R_x$ meters in the $x$-axis). In indoor scenes, the $x$ and $y$ axes represent the horizontal room dimensions, thus being inherently bounded, while the $z$ axis represents the vertical room height, inherently bounded by the floor-to-ceiling distance. In outdoor scenarios (e.g., autonomous driving), the $x$ and $y$ axes correspond to the forward and lateral directions, bounded by the sensor's capture range (horizontal field of view). For example, in Waymo dataset ~\cite{DBLP:journals/corr/abs-1912-04838}, the horizontal LiDAR radius is $\sim$75 meters, yielding $R_x, R_y\leq 150$ meters. The vertical $z$ axis range $R_z$ is typically cropped to a few meters~\cite{yin2021center} to focus on the height of relevant nearby objects. 
Since raw point cloud coordinates are bounded by a spatial range, the corresponding quantized integer voxel coordinates are also bounded by spatial range $\left(\left\lfloor \frac{R_x}{g_x} \right\rfloor,\; \left\lfloor \frac{R_y}{g_y} \right\rfloor,\; \left\lfloor \frac{R_z}{g_z} \right\rfloor\right)$, where $g = (g_x, g_y, g_z)$ is the grid size vector, resulting in a finite, well-defined 3D grid whose exact extent depends on the application.

\begin{figure}[h]  

    \centering  
    \includegraphics[width=\linewidth]{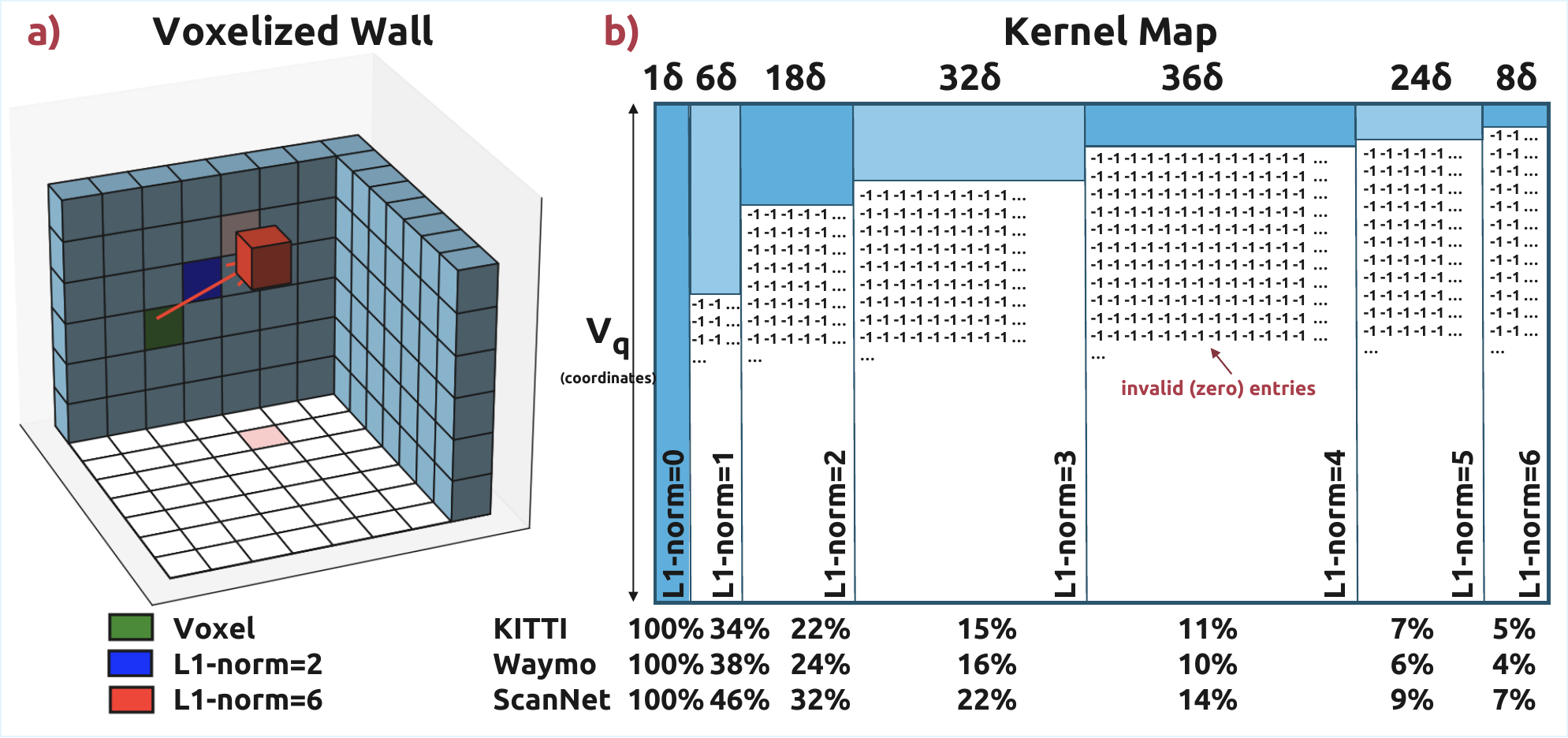}
    \vspace{-20pt}
    \caption{a) A voxelized wall surface. b) In a submanifold layer with $K$$=$5, the average density of kernel map columns for weight offsets grouped by their L1-norm for three different datasets.}
    \label{fig:motivation-lnorm-prop}
\end{figure}

\noindent\camera{\textbf{(3) Neighboring Property:}} 
\camera{\emph{Voxel coordinates belonging to the same object or surface are likely to differ by only small offsets across each spatial dimension.} Voxel points capture \textbf{continuous object surfaces} rather than randomly scattered points. When a voxel is occupied (i.e., non-zero) at position ($v_x,v_y,v_z$) on a voxelized object surface, nearby voxels at small displacements, measured by the L1-norm\footnote{\noindent L1 norm of weight offset $\delta = (\delta_x, \delta_y, \delta_z)$ is  $|\delta_x| + |\delta_y| + |\delta_z|$.}, are highly likely to also be occupied (i.e., be non-zero) and belong to the \emph{same} surface. In contrast, voxels at large L1-norm displacements, where coordinates shift across multiple axes simultaneously (e.g.,  along $x$, $y$, and $z$) by larger displacements, are likely to fall in empty space.} For example, ~\autoref{fig:motivation-lnorm-prop}a shows a voxelized wall with a green voxel (cube) annotated. The blue voxel that has been shifted by (0,1,1) (small L1-norm of 2) from the green voxel, belongs to the same wall surface, while the red voxel, shifted by (2,2,2) (larger L1-norm of 6), corresponds to empty space (air).

This geometric property \camera{has the following important implication} for submanifold layers, which constitute over 70\% of layers in state-of-the-art point cloud networks, where output coordinates are identical to input coordinates.
\emph{In submanifold convolution, kernel map columns for weight offsets with smaller L1 norms consistently exhibit higher density than those with larger L1 norms, reflecting the geometric continuity of object surfaces captured by voxel data. }

We comprehensively examine \emph{hundreds} of voxel point clouds from publicly available indoor and outdoor datasets and observe that in submanifold layers, kernel map columns for weight offsets with small L1-norms consistently exhibit higher density than columns for larger L1-norms.  This occurs because \camera{the weight offsets are associated with coordinate displacements, and voxel coordinates belonging to the same object surface tend to differ by small L1-norm displacements, while coordinates differing by larger L1-norms typically correspond to empty space.} 
In a \SpConv layer with input stride $s_p$, the L1-norm of weight offsets takes values from $0$ up to L1NormMax $=\frac{3(K-1)}{2} \times s_p$, in steps of $s_p$.
\autoref{fig:motivation-lnorm-prop}b shows the average density of kernel map columns for weight offsets grouped by L1 norm (0 to 6) across three datasets in a layer with \camera{kernel size} $K$$=$5 (125 total weight offsets) and \cg{$s_p$=$1$}. The column for weight offset (0,0,0) exhibits 100\% density, because by definition of submanifold convolution, each output voxel coordinate maps to itself as an input coordinate. 
Columns for weight offsets with L1-norm of 1 average 39.4\% density across datasets, while those with L1-norm of 6 average only 5.4\% density.

%% file: content/mechanism.tex
\section{\SysName Design}

\subsection{Overview}

\SysName is the first voxel-property-aware \SpConv engine  that integrates fast mapping search, reduces coordinate access costs, improves parallelism in voxel indexing, and supports flexible hybrid dataflow processing.

\SysName \camera{consists of the following} four key ideas:\\
\noindent\textbf{1. One-Shot Z-Delta Search Mapping.} We design an one-shot search algorithm that builds the kernel map with \emph{no} pre-processing phase. By intelligently partitioning weight offsets into $K^2$ groups of $K$ offsets each, 
we perform only one binary search per group to resolve the first query, then we use fast localized search to resolve the remaining $K$$-$1 queries of that group. This approach significantly reduces memory accesses and computational cost. \\
\noindent\textbf{2. Packed-Native Voxel Indexing.} Leveraging the bounded property, we pack three coordinate values into a single (32-bit or 64-bit) integer, and propose packed-native voxel indexing that \emph{directly} processes packed data, completely eliminating unpacking/repacking and reducing data access costs. \\
\noindent\textbf{3. Adaptive Hybrid-Dataflow Feature Computation.}
We exploit the \camera{neighboring} property in feature computation and propose a flexible hybrid dataflow scheme where different weight offsets in kernel map can be processed with either output- or weight-stationary way. 
This approach spans the full dataflow range, from single-dataflow executions to flexible hybrid combinations. In this way, we find the optimal trade-off between unnecessary computations and expensive atomic operations for each layer's specific configuration. \\
\noindent\textbf{4. Network-Wide Voxel Indexing.} Via detailed network analysis,  we find voxel indexing kernels have \emph{no} dependencies, thus we execute them concurrently at the network start across multiple GPU SMs to improve execution parallelism.

\subsection{One-Shot Z-Delta Search Mapping}\label{sec:mechanism-zdelta}

The \textbf{goal} of our mapping search scheme is to achieve (i) minimal pre-processing overhead and (ii) fast search lookup operations for the queries $\mathbf{q_i} + \boldsymbol{\delta}_\mathbf{k}$  ($\mathbf{q_i}$ output coordinate and $\boldsymbol{\delta}_\mathbf{k}$  weight offset).  
\SysName's algorithm \emph{completely} eliminates pre-processing leveraging the following key observation.

\noindent\textbf{Key Observation}: \emph{When input voxel coordinates of the first \SpConv layer of a network are lexicographically sorted, \textbf{all} \SpConv layers maintain \textbf{sorted} coordinates throughout the network.} 
In submanifold layers ($s_l = 1$), output coordinates are identical to input coordinates, directly preserving the sorted order. In downsampling layers ($s_l > 1$), the fastest downsampling schemes of state-of-the-art \SpConv engines extract  \emph{unique} output coordinates by rounding input coordinates, and \emph{sorting} the rounded results to remove duplicates, i.e., a process that inherently produces \emph{sorted} output coordinates $V_q$. Since each layer's output become the next layer's input, sorting propagates naturally through the network. Thus, sorting coordinates once at the first layer guarantees sorted coordinates throughout all layers.

Our \textbf{one-shot} search algorithm eliminates the pre-processing by operating directly on the default coordinate arrangement: input and output coordinates of all network layers stored in separate sorted arrays, requiring only a single sort in the first layer's input coordinates. To search for queries $\mathbf{q_i} + \boldsymbol{\delta}_\mathbf{k}$ in each layer's input coordinate array,  we generate these queries \emph{on-the-fly}  by adding the current weight offset $\boldsymbol{\delta}_\mathbf{k}$ to each sorted output coordinate $\mathbf{q_i}$.

A simple binary search algorithm can exploit this sorted arrangement without pre-processing: parallel threads generate queries $\mathbf{q_i} + \boldsymbol{\delta}_\mathbf{k}$ on-the-fly and perform binary search on the input coordinate array to find mappings between input coordinates, output coordinates and weight offsets.
This simple algorithm performs $|V_q| \times K^3 $ independent binary searches—one per generated query. 
However, despite exhibiting high temporal data locality when searching \emph{sorted} queries~\cite{Yang2024Minuet}, performing one full binary search per query in the entire input coordinate array in global memory remains expensive due to the large number of memory accesses and high computational \camera{cost}.

\begin{figure}[h]
    \centering 
    \includegraphics[width=\linewidth]{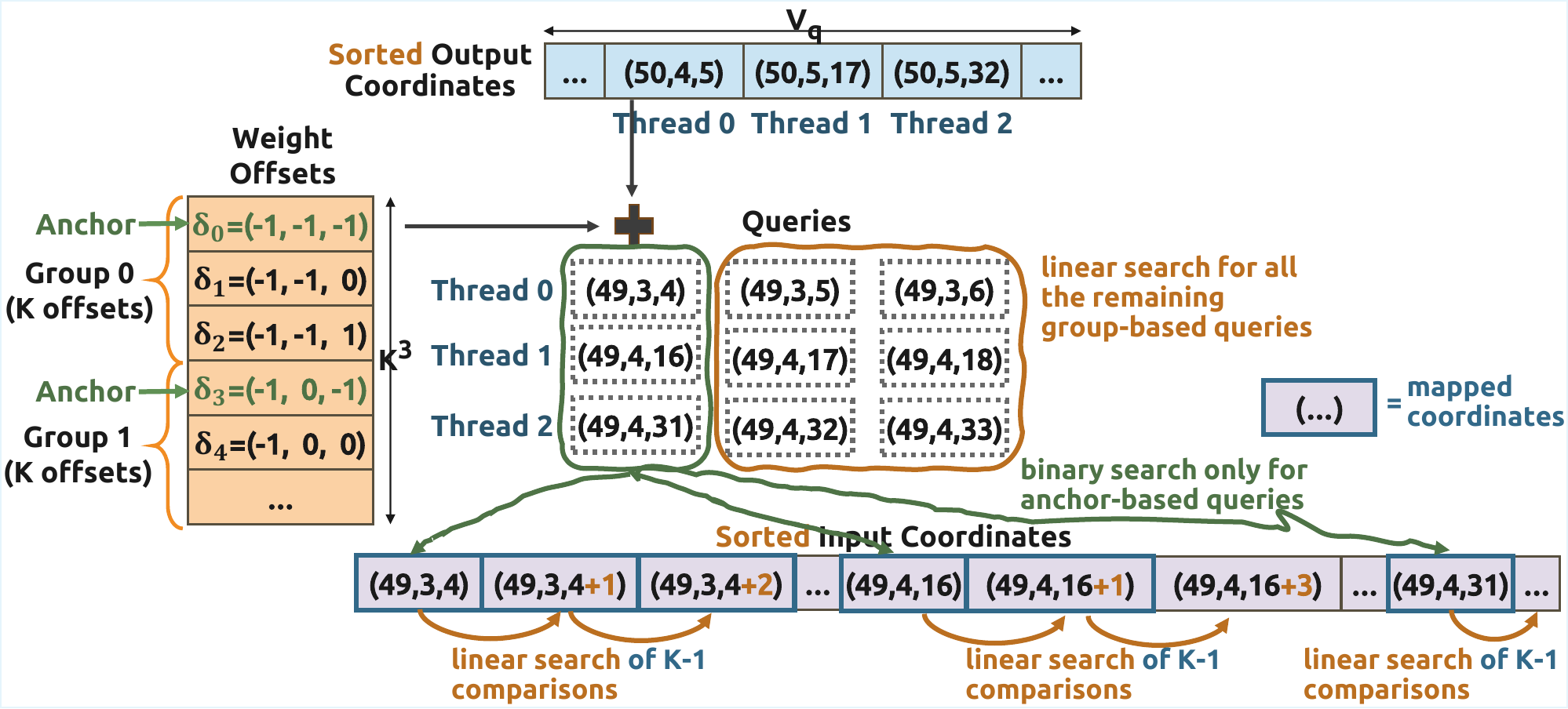}
    \vspace{-22pt}
    \caption{\SysName's one-shot z-delta search algorithm.}
    \label{fig:mechanism-z-delta-search}
\end{figure}

\SysName leverages the \textbf{Integer Property} of voxel data to reduce the computational \camera{cost}, while maintaining a one-shot  design without pre-processing.
\autoref{fig:mechanism-z-delta-search} presents our proposed \textbf{z-delta search} algorithm in a layer with kernel size $K$=3 and input stride $s_p$=1. We group the $K^3$ weight offsets $\boldsymbol{\delta}$ in $K^2$ groups of $K$ offsets each, where offsets within a group have \emph{identical}  $x$ and $y$ values and consecutive $z$ values, e.g., the offsets $(-1, -1, -1), (-1, -1, 0), (-1, -1, 1)$ form group 0.
We distribute the sorted output coordinates and weight offset groups among parallel threads: each thread  processes all the $K$  queries from  $\mathbf{q_i} + \boldsymbol{\delta}_\mathbf{j}$ to  $\mathbf{q_i} + \boldsymbol{\delta}_\mathbf{j+K-1}$ for a single coordinate $\mathbf{q_i}$ and weight offset group.
This grouping ensures the $K$ query coordinates range from $(x, y, z)$ to $(x, y, z + (K$-$1) \times s_p)$—sharing identical $x$ and $y$ values and having consecutive  $z$ values increasing by $s_p$.
For instance, for output coordinate (50,4,5), thread 0 processes weight offset group 0 and generates $K$=3 queries (49,3,4), (49,3,5) and (49,3,6) with identical $x$ and $y$ values and $z$ values increasing by $s_p$=1.

\noindent\textbf{Key Observation:} \emph{For a thread processing the $K$ queries  from $(x, y, z)$ to $(x, y, z + (K$-$1) \times s_p)$ for the same weight offset group, the \textbf{sorted} and \textbf{integer-based} nature of the input coordinate array ensures that consecutive queries $(x, y, z)$ and $(x, y, z + s_p)$, if they exist, must occupy consecutive positions in the \camera{sorted input} array. 
} 
There cannot be three coordinates $(x, y, z), (x, y, z'), (x, y, z+s_p)$ where $z<z'<z+s_p$, in the input coordinate array. 
This follows directly from the \textbf{Integer Property} and \SpConv definition: since input coordinates are integer  and multiples of stride $s_p$ and  are lexicographically sorted in input array, \textbf{no} coordinate can exist between $(x, y, z)$ and $(x, y, z+s_p)$, with the identical $x$ and $y$ values (input coordinates must differ by some multiples of $s_p$).


Our z-delta search algorithm (\autoref{fig:mechanism-z-delta-search}) exploits this key observation to reduce the computational \camera{cost}, and improve \camera{data} locality.
For $|V_q|$ output coordinates and $K^3$ weight offsets, z-delta search performs only $|V_q| \times K^2$ binary searches \camera{(instead of $|V_q| \times K^3$)}.
For each weight offset group, the offset with the smallest z-value serves as the \textbf{anchor} (e.g., $\delta_0=(-1,-1,-1)$ is the anchor in group 0). For the $K$ queries generated from the same weight offset group, we perform binary search in the input coordinate array \textbf{only} for the anchor-based query (e.g., (49,3,4) query for group 0). The binary search returns the position the matched input coordinate—or, if no match exists, the next larger coordinate. 
We resolve the remaining $K$$-$1 queries (e.g., (49,3,5) and (49,3,6) queries for group 0) via a quick localized linear search starting from this position: the local linear search checks at most $K$$-$1 consecutive positions in the input array to look for potential matchings for the remaining $K$$-$1 queries.
Since these positions are contiguous in memory, the localized linear search achieves optimal cache efficiency with minimal data access latency. 
Overall, our z-delta search algorithm enables one-shot kernel map construction with low computational cost and high data locality, significantly reducing expensive global memory accesses.

\subsection{Packed-Native Voxel Indexing}\label{sec:mechanism-packing}

Prior \SpConv engines use 32 bits for \emph{each} of the three voxel coordinates of a triplet. Instead, \SysName leverages the \textbf{Bounded Property} to represent each integer coordinate with fewer bits and pack all three coordinates into a \emph{single} integer value. 

For many real-world applications (e.g., robotics, automotive), packing all three coordinates into a single 32-bit value is sufficient. \da{For example, in outdoor scenes captured by LiDAR sensors with horizontal ranges $R_x$ and $R_y$ of up to 400 meters and a vertical range $R_z$ of up to 20 meters, when quantizing the coordinates with a grid size of 0.1 meters,  coordinates can be represented using 12, 12, and 8 bits for $v_x$, $v_y$, and $v_z$, respectively.}
For highly demanding applications, \SysName can  pack voxel coordinates into a single 64-bit value (See \autoref{fig:end-to-end-3090}). 
Note that 64-bit packing for the voxel coordinate triplet is sufficient for all current real-world \SpConv applications: 64-bit packing can represent scenes with ranges $R_x$, $R_y$, $R_z$ up to several kilometers and satisfying spatial resolutions of the grid size vector $g$ up to millimeters, which provides the maximum accuracy that today's 3D sensors can support to capture raw point cloud data. 
\SysName packs the triplet ($v_x$, $v_y$, $v_z$) as one single value: each $v_i$ integer value is within an integer range of $\lfloor \frac{R_i}{g_i} \rfloor$, thus the necessary bits to express it are at most $b_i = \log(\lfloor \frac{R_i}{g_i}  \rfloor)$.
The first $b_x$ most significant bits \camera{is} for $x$, \camera{the} next $b_y$ bits for $y$ and the remaining $b_z$ bits for $z$.
In §\ref{sec:evaluation}, we use 12, 12, and 8 bits for $v_x$, $v_y$ and $v_z$, respectively. However, the bit allocation is flexible and can be easily configured. 


The \textbf{key challenge} with packed coordinates is how to avoid unpacking and repacking operations in the \SpConv voxel indexing kernels (feature computation does not use voxel coordinates).
\SysName enables \textbf{packed-native voxel indexing} in both downsampling and mapping kernels,  \emph{completely} eliminating any unpacking/repacking, thus the only packing required for voxel data is performed once \cg{in the initial coordinates.} 

In downsampling, rounding each packed input coordinate with the output stride $s_q$=$2^m$ can be efficiently performed by bitwise adding two 32-bit values: the packed coordinate $packed(p_j)$ 
and the 
$mask =
\underbrace{11\cdots 1}_{b_x - m}%
\underbrace{00\cdots 0}_{m} \;
\underbrace{11\cdots 1}_{b_y - m}%
\underbrace{00\cdots 0}_{m} \;
\underbrace{11\cdots 1}_{b_z - m}%
\underbrace{00\cdots 0}_{m}
$.
Downsampling also requires sorting packed coordinates to remove duplicates. Notably, sorting \emph{directly} in packed format inherently preserves lexicographic order: $p_i > p_j \Leftrightarrow packed(p_i) > packed(p_j)$. Therefore, we can directly sort the packed coordinates without unpacking.

In mapping, we apply to the output coordinates $q_i$ and weight offsets $\delta_k$ exactly the same packing format as used in input coordinates, representing each as a single packed value. We leverage the property that $packed(q_i) + packed(\delta_k) = packed(q_i + \delta_k)$, which allows us to directly generate queries ${q_i} + {\delta}_{k}$ as $packed(q_i) + packed(\delta_k)$ represented in the same packed format. This eliminates the need to unpack the $packed(q_i)$, add weight offset to generate $q_i + \delta_k$, and repack the result. The lookup searches of mapping can be also directly performed using packed queries,   
as a packed format  preserves lexicographic order: $(q_i + \delta_k) > p_j \Leftrightarrow packed(q_i) + packed(\delta_k) > packed(p_j)$.

\SysName's packed-native voxel indexing provides four performance benefits over prior works: (i) the memory footprint for voxel coordinates is lower (e.g., 3$\times$ lower for 32-bit packing), (ii) memory reads and writes for voxel coordinates are faster thanks to 
fewer global memory accesses, (iii) lexicographic sorting during downsampling is faster when sorting a single (32-bit or 64-bit) value compared to sorting three separate (32-bit) values, and (iv) lexicographic comparisons for lookup operations during mapping are faster—a single comparison of a packed coordinate value is faster than three separate comparisons of individual coordinates.

\subsection{Adaptive Hybrid-Dataflow Feature Computation}\label{sec:mechanism-dataflow}

Output-stationary and weight-stationary dataflows each perform best under different kernel map densities and \SpConv layer characteristics. When the kernel map is relatively dense (fewer invalid entries), output-stationary dataflow performs better by avoiding kernel map filtering and eliminating expensive atomic operations during feature computation. Instead, when the kernel map is relatively sparse (large number of invalid entries), weight-stationary dataflow is more efficient by skipping numerous zero-valued multiplications.

In submanifold layers, which typically constitute  $\sim$70\% of \SpConv layers in state-of-the-art point cloud networks, the \textbf{Neighboring Property} states that kernel map columns corresponding to weight offsets with smaller L1 norms consistently exhibit higher density than those with larger L1 norms. For example, in \SpConv layers with $K\geq 5$, most weight offset columns  are up to 10$\times$ sparser than a small subset of significantly denser columns.
This property implies that using a single dataflow (either output-stationary or weight-stationary) for feature computation is suboptimal. 

\begin{figure}[h!]  
    \centering 
    \includegraphics[width=0.99\linewidth]{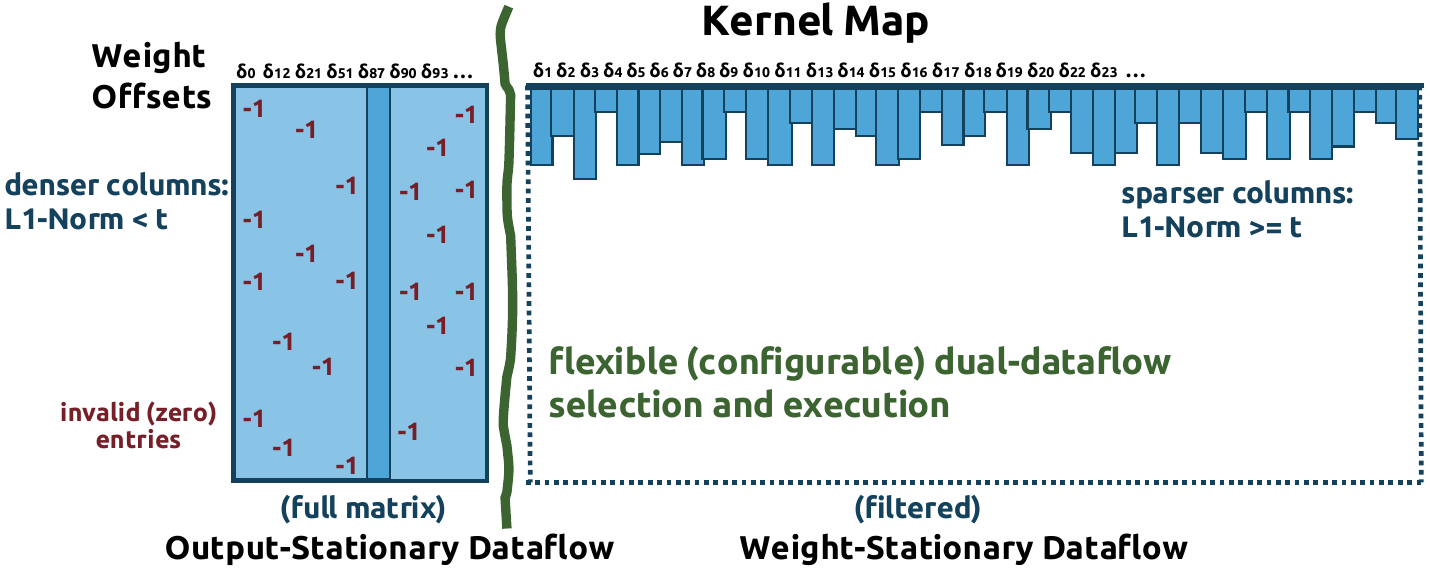}
    \vspace{-10pt}
    \caption{\SysName's adaptive hybrid-dataflow feature computation.}
    \label{fig:mechanism-dual-dataflow}
\end{figure}

To address this, \SysName's feature computation scheme (i) supports the full dataflow spectrum, including single-dataflow execution (all weight offsets processed using one dataflow) and hybrid dual-dataflow execution (different weight offset subsets processed using different dataflows), and (ii) minimizes post-processing costs for kernel map rearrangement required by each dataflow.
\autoref{fig:mechanism-dual-dataflow} provides an overview of \SysName's adaptive hybrid dataflow for feature computation.

We track the density of all kernel map weight offsets, 
and select the appropriate dataflow for each weight offset by classifying them as \emph{dense} or \emph{sparse} using a configurable threshold $t$:  
weight offsets with L1-norm $< t$ are are classified as \emph{dense}, and processed with output-stationary dataflow, and weight offsets with L1-norm $\geq t$ are classified as \emph{sparse}, and processed with weight-stationary dataflow. When \cg{$t$=L1NormMax+1}
, all columns are considered dense, degenerating to full output-stationary execution. When $t$=$0$, all columns are considered sparse, degenerating to full weight-stationary execution.

\SysName tunes the threshold value $t$ for each \SpConv layer  using the same general tuning approach as prior state-of-the-art works~\cite{Tang2023TorchSparse++, Yang2024Minuet}\camera{, that also have tunable feature computation schemes}. 
We sample a few point clouds from the dataset and  measure  kernel map build time and feature computation latency for different (integer) $t$ values, then select the  $t$ value that minimizes total latency. \camera{In §\ref{sec:evaluation}, we show that our tuning  scheme \da{can yield threshold selections for the layers of our evaluated networks that can provide high performance in inference} across datasets of different patterns}. Note that this is a \emph{one-time} tuning step performed only once before inference having negligible overhead. \camera{For example, using 5 samples, \SysName's tuning time for ResN network that has 21 \SpConv layers (See §\ref{sec:methodology}) is less than 3 seconds. Prior state-of-the-art works~\cite{Tang2023TorchSparse++, Yang2024Minuet} exhibit for their respective tuning steps a comparable time of 2-5 seconds. Hence, if re-tuning is required, \da{to accommodate a scenario where the density of the scenes may substantially change over time (e.g., transitioning from sparse highway scenes to high-density urban environments), re-tuning} can be effectively performed without significant latency costs.}

The \textbf{key challenge} \camera{in feature computation step} is how to fully support adaptive hybrid dataflow  executions with \emph{low} post-processing overheads in voxel indexing. 
Output-stationary dataflow requires the kernel map to be in $|V_q| \times K^3$ layout, while weight-stationary requires it transposed (§\ref{sec:background-execution-steps}) to $K^3 \times |V_q|$ layout (also filtered from invalid indices). These layouts enable \textbf{memory coalescing} in kernel map construction: adjacent threads within a thread block write to adjacent memory locations, maximizing memory bandwidth utilization. 
Similarly, hybrid dual-dataflow with $K_{dense}$ \emph{dense} weight offsets and $K_{sparse}$ \emph{sparse} weight offsets, where $K_{dense}$+$K_{sparse}$=$K^3$, requires two kernel map submatrices: one of  $|V_q| \times K_{dense}$ layout for output-stationary and another of $K_{sparse} \times |V_q|$ layout (also filtered) for weight-stationary, both enabling memory coalescing. These different layout requirements could incur high post-processing overheads in voxel indexing, as the entire kernel map must be rearranged into the appropriate layout for each dataflow, thus potentially negating the performance benefits of hybrid dual-dataflow execution. 

\begin{figure}[h]
    \centering 
    \includegraphics[width=\linewidth]{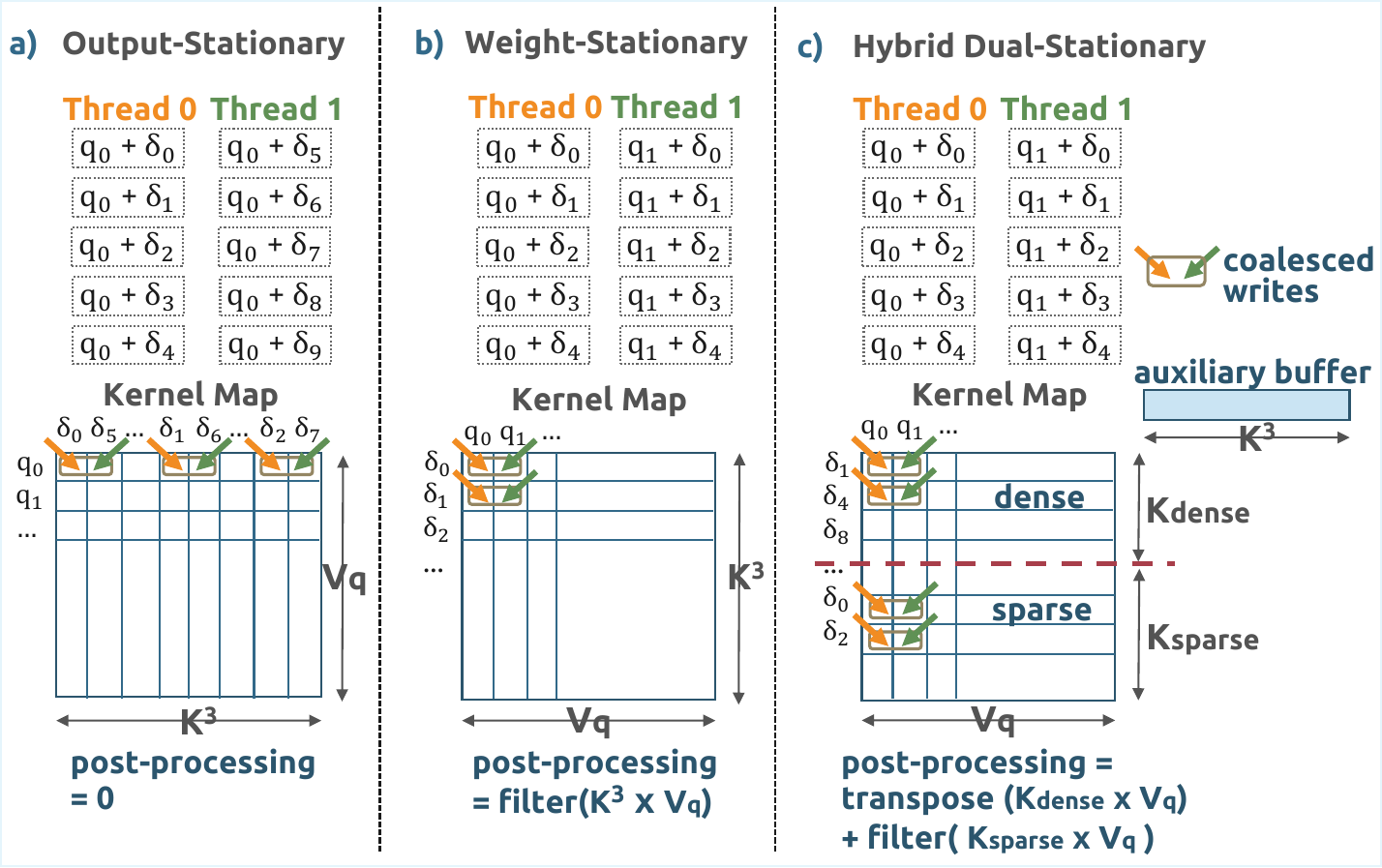}
    \vspace{-26pt}
    \caption{\SysName's post-processing in three dataflow scenarios.}
    \label{fig:mechanism-dataflow-coal}

\end{figure}

\SysName achieves low post-processing time across all three dataflow scenarios (\autoref{fig:mechanism-dataflow-coal}) by optimizing how queries are distributed across parallel threads within a thread block.
I) Full output-stationary 
\cg{$t$=L1NormMax+1} (\autoref{fig:mechanism-dataflow-coal}a): \SysName has \textbf{no} post-processing (zero). Our z-delta search distributes different weight offsets $\boldsymbol{\delta}_\mathbf{j}$ for the same output coordinate $\mathbf{q_i}$ across multiple threads  within a thread block, enabling coalesced memory writes that directly produce the required $|V_q| \times K^3$ layout. 
II) Full weight-stationary $t$=$0$ (\autoref{fig:mechanism-dataflow-coal}b): \SysName has post-processing \emph{only} for filtering invalid entries (no transposition is needed). Our z-delta search distributes different output coordinates $\mathbf{q_i}$ for the same weight offset $\boldsymbol{\delta}_\mathbf{j}$ across threads within a thread block, enabling coalesced memory writes that directly produce the required $K^3 \times |V_q|$ layout. 
III) Hybrid dual-dataflow $0 < t \leq$ \cg{L1NormMax, where $t$ is multiple of $s_p$}  (\autoref{fig:mechanism-dataflow-coal}c): Our z-delta follows the weight-stationary parallelism scheme (distributing different output coordinates $\mathbf{q_i}$ for the same weight offset $\boldsymbol{\delta}_\mathbf{j}$ across threads \camera{within a block}). We employ an auxiliary buffer  of size $K^3$ to help us classify weight offsets as \emph{dense} and \emph{sparse}  and  partition the kernel map into two submatrices: one with $K_{dense} \times |V_q|$ layout storing all dense weight offset rows contiguously, and another with $K_{sparse} \times |V_q|$ layout storing all sparse weight offset rows contiguously.  Post-processing then transposes \emph{only} the dense submatrix to produce the required $|V_q| \times K_{dense}$ layout, and filters \emph{only} the sparse submatrix to remove invalid entries.
The post-processing time in hybrid dual-dataflow is minimal and comparable to that of the full weight-stationary dataflow.

In submanifold layers, \SysName further reduces post-processing time for weight-stationary dataflow and weight-stationary submatrix in hybrid dual-dataflow by exploiting a symmetry property from prior work~\cite{tang2022torchsparse}: weight offsets $\delta_l=(b,c,d)$ and $\delta_n=(-b,-c,-d)$ are symmetric, i.e., if $M[i, l]=j >0$ in the kernel map, then $M[j, n]=i>0$. 
Weight-stationary feature computation, where each thread block operates on a fixed weight offset, 
compute outputs features for symmetric weight offset pairs by reading only the first 
half of the kernel map. Thus, only half of kernel map is stored and filtered,  reducing post-processing time.

\subsection{Network-Wide Voxel Indexing}\label{sec:mechanism-network-voxel-indexing}

We carefully examine point cloud network operators and make the following \textbf{Key Observation}: \emph{The voxel indexing step of a \SpConv layer has \textbf{no} true data dependencies with voxel indexing or feature computation steps of other layers}. Our profiling also shows that voxel indexing incurs low GPU SM utilization, as it performs only lightweight comparisons. 

Voxel indexing step includes downsampling to generate output coordinates, and  mapping, that uses them to construct the kernel map needed in feature computation step. Typical downsampling for layer $i$ is performed recursively as $V_i$ =  $\lfloor \frac{V_{i-1}}{2^i} \rfloor \times 2^i $  using the voxel coordinates $V_{i-1}$ from the previous downsampling layer. However, we find that this recursive formula can be transformed into a closed-form expression using \emph{only the initial input} coordinates $V_0$:
{\setlength{\abovedisplayskip}{2pt}
\setlength{\belowdisplayskip}{2pt}
\begin{align*}
V_i &= \left\lfloor \frac{V_{i-1}}{2^i} \right\rfloor 2^i
= \left\lfloor 
    \frac{\left\lfloor \frac{V_{i-2}}{2^{i-1}} \right\rfloor 
          2^{i-1}}{2^{i}} 
  \right\rfloor 2^i \nonumber \\
&= \left\lfloor \frac{V_{i-2}}{2^i} \right\rfloor 2^i
= \cdots 
= \left\lfloor \frac{V_0}{2^i} \right\rfloor 2^i
\end{align*}
}

Thus, downsampling kernels of different layers have \emph{no} true dependencies with each other: voxel coordinates of layer $i$ can be \emph{directly} extracted from the initial coordinates $V_0$. Moreover, mapping kernels are also mutually \emph{independent}, since each mapping kernel only requires its associated layer's downsampled coordinates to build the kernel map consumed by  subsequent feature computation.

\SysName enables \textbf{network-wide voxel indexing} by executing voxel indexing steps for \emph{all} layers in parallel at the network start via a two-phase process. First, all downsampling kernels across all layers  are executed \emph{concurrently} using CUDA streams distributed across multiple GPU SMs, generating coordinate sets for their corresponding mapping kernels. Second, all mapping kernels are executed \emph{concurrently}, again using streams across multiple SMs to produce kernel maps for all layers. These kernel maps are stored in global memory and consumed by their corresponding feature computation steps as inference progresses. 

\begin{figure}[h]  
    \centering 
    \includegraphics[width=\linewidth]{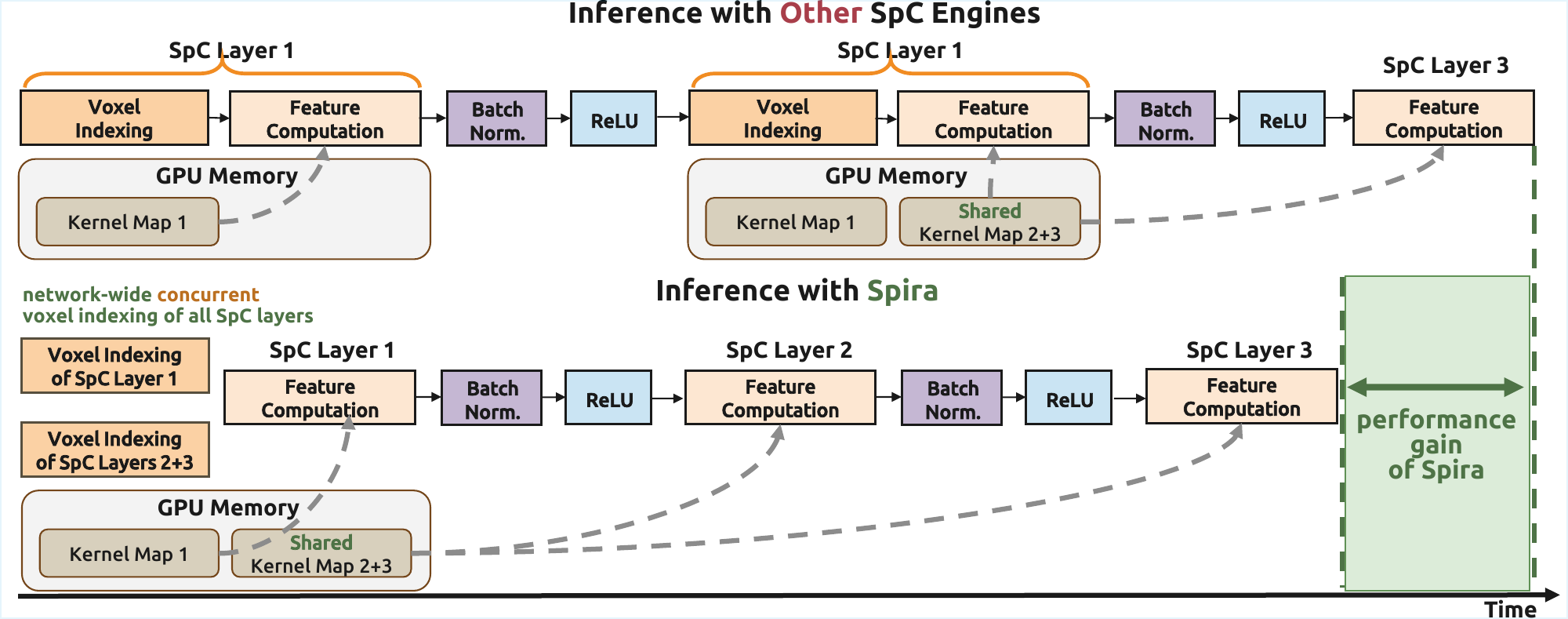}
    \vspace{-24pt}
    \caption{\camera{Inference of existing \SpConv engines versus network-wide voxel indexing inference of \SysName.}}
    \label{fig:mechanism-merged-indexing}
\end{figure}

\camera{In a typical network, each \SpConv layer does not require a unique kernel map, instead multiple \SpConv layers with same key characteristics (i.e., layers with same input stride, layer stride and kernel size) have identical kernel maps. Consequently, some kernel maps can be shared and re-used across multiple \SpConv layers.  For instance, consecutive submanifold layers with the same kernel size share the same kernel map. In \autoref{fig:mechanism-merged-indexing}, the kernel map of \SpConv layer 2 is re-used in \SpConv layer 3.  
As a result, existing \SpConv engines ~\cite{Choy2019Minkownski,Tang2023TorchSparse++,Yang2024Minuet} maintain the kernel maps in memory during inference and reuse them in multiple layers. The key difference of \SysName versus prior existing \SpConv engines is when kernel maps are computed, depicted in \autoref{fig:mechanism-merged-indexing}. Prior engines compute and store kernel maps progressively, as layers are executed during inference, while \SysName computes all kernel maps upfront (at the start of inference). Therefore, both approaches converge to have similar memory footprints for the kernel maps, as well as for all other inference data, including features and weights which are the same across engines. The memory footprint of the kernel maps is on average $\sim$40MB stored in global memory—affordable even for edge GPUs.}
\vspace{-2pt}

\vspace{-2pt}

\camera{In submanifold layers with weight-stationary dataflow, we only store \emph{half} the kernel map thanks to exploiting the symmetry property explained in §\ref{sec:mechanism-dataflow}, thus} further reducing the memory footprint.
\SysName's network-wide voxel indexing improves execution parallelism and GPU utilization.

%% file: content/evaluation.tex
\section{Evaluation}\label{sec:evaluation}

\subsection{Evaluation Methodology}\label{sec:methodology}

\noindent\textbf{Hardware Systems / GPUs.} \camera{We evaluate \SysName on six NVIDIA GPUs spanning consumer, workstation, datacenter, and edge categories: the consumer GPUs RTX 3090 (24GB) and GTX 1060 (6GB); the workstation GPU Quadro RTX 5000 (16GB); the datacenter GPUs Tesla A100 (80GB) and H100 (80GB); and an edge GPU platform widely used in robotics and edge applications, the Jetson Orin AGX (64GB shared), operated under a 50W power budget. Unless otherwise stated, all detailed evaluation results are reported on the RTX 3090 GPU.}

\noindent\textbf{Neural Networks.} We evaluate three 
point cloud networks: SparseResNet (21 layers) (\textbf{ResN}), MinkUNet (42 layers) (\textbf{UNet}) ~\cite{Choy2019Minkownski}, and Centerpoint Large (20 layers) (\textbf{ResNL}) ~\cite{chen2023voxelnextfullysparsevoxelnet}, 
\cg{which uses \SpConv layers with $K$=5 (instead of 3) in all stages of ResNet backbone}.

\noindent\textbf{Datasets.} We evaluate 
three large-scale point cloud datasets: SemanticKITTI (\textbf{KITTI}) ~\cite{DBLP:journals/corr/abs-1904-01416}, which includes outdoor LiDAR scans for self-driving scenarios \camera{(average scene density: 0.11\%)}, \textbf{ScanNet} ~\cite{dai2017scannetrichlyannotated3dreconstructions}, which contains RGB-D scans of indoor environments \camera{(average scene density: 1.57\%)}, and \textbf{Waymo} ~\cite{DBLP:journals/corr/abs-1912-04838}, which provides large-scale outdoor scenes for automotive object detection \camera{(average scene density: 0.16\%)}. 

\noindent\textbf{Comparison Points.} We compare \SysName with two state-of-the-art \SpConv engines: (i) Minuet and (ii) TorchSparse++. In \SysName, we account for both sorting and packing the initial input coordinates. We do not include tuning time for any engine as tuning  happens only once and is before inference. 
\camera{Unless otherwise stated, we use 16-bit float precision in our experiments and a batch size of 1, because inference workloads for point cloud networks are typically latency-oriented rather than throughput-oriented}. 

\subsection{End-to-End Inference Performance}\label{sec:end_to_end}

\begin{figure}[h]  
    \centering 
    \includegraphics[width=\linewidth]{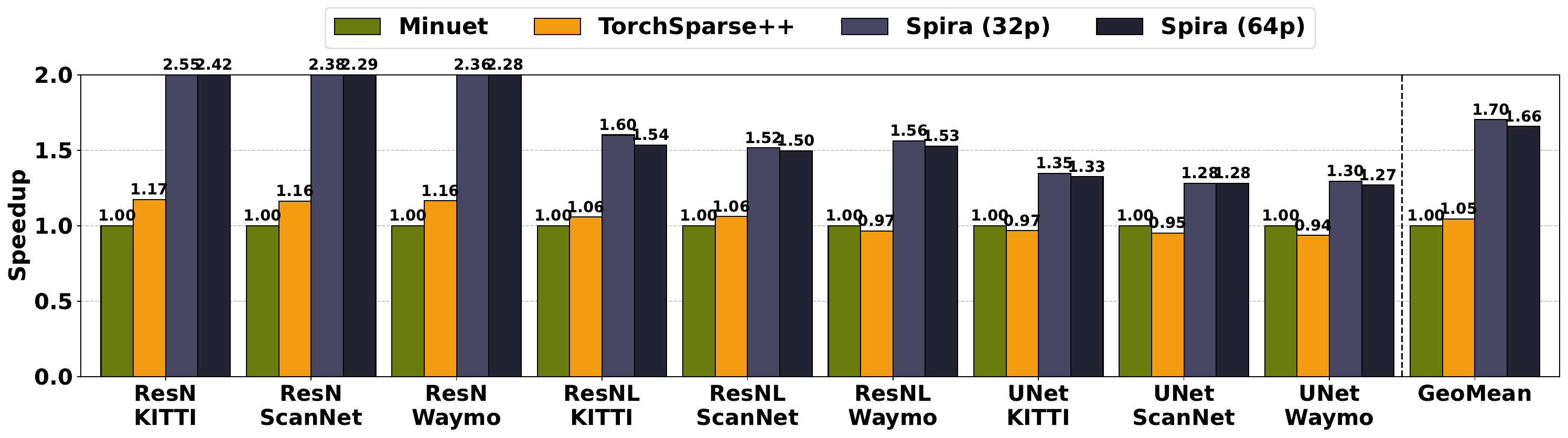}
    \vspace{-22pt}
    \caption{End-to-end inference performance of all \SpConv engines using various point cloud networks and datasets on RTX 3090.}
    \label{fig:end-to-end-3090}

\end{figure}

\begin{figure}[h]  
    \centering 
    \includegraphics[width=\linewidth]{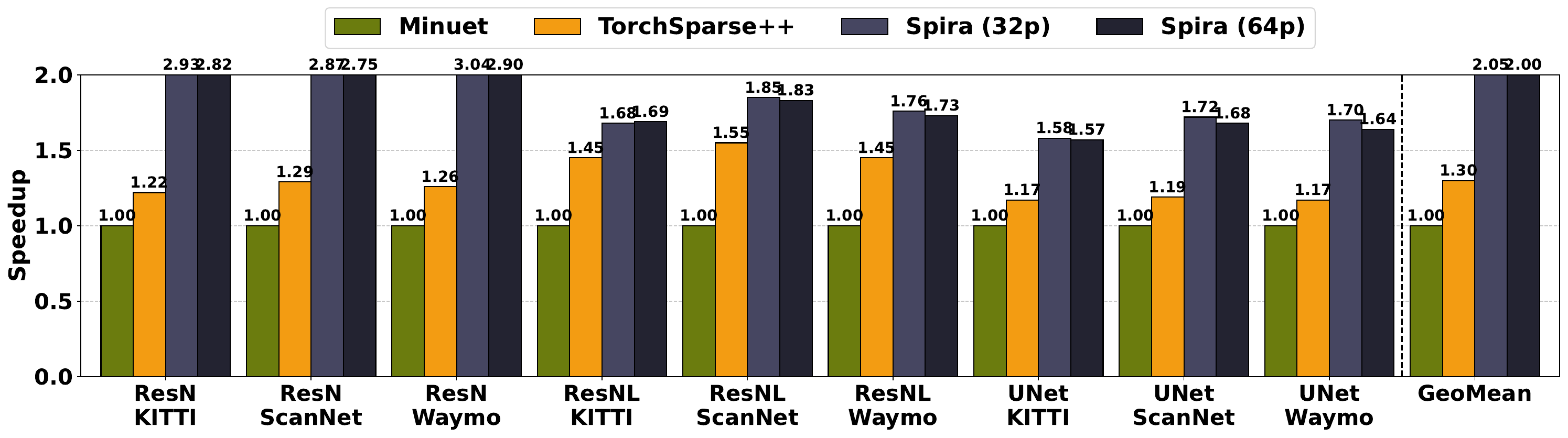}
    \vspace{-22pt}
    \caption{\camera{End-to-end inference performance of all \SpConv engines using various point cloud networks and datasets on A100.}}
    \label{fig:end-to-end-A100}

\end{figure}

\camera{\autoref{fig:end-to-end-3090} and~\autoref{fig:end-to-end-A100}} show end-to-end inference performance of all engines on RTX 3090 \camera{and A100}, respectively, across different networks and datasets. \camera{See also Appendix §\ref{sec:more_gpus} for detailed results on the remaining GPUs}. 
\SysName uses  32-bit or 64-bit packed-native voxel indexing. 

We make three key observations. First, \camera{TorchSparse++ outperforms Minuet on A100, but achieves comparable performance on RTX 3090. This is because on the A100 GPU, the output-stationary dataflow performs better than the weight-stationary in the vast majority of layers across all networks, and Minuet does not support output-stationary. This trend is attributed to the A100’s higher compute throughput and memory bandwidth, which mitigate the cost of redundant computations. In contrast, on the RTX 3090, the weight-stationary dataflow outperforms the output-stationary in approximately half of the layers within the UNet and ResNL networks. Second, across all evaluated networks and datasets, Spira consistently delivers the best performance over prior \SpConv engines on both RTX 3090 and A100, achieving average speedups of \SpiraEndToEndAvgThirty and \SpiraEndToEndAvgAHundred over Minuet, and of $1.62\times$ and $1.58\times$  over TorchSparse++, on RTX 3090 and A100, respectively.}
Third, \SysName with 64-bit coordinate packing versus 32-bit of has negligible performance impact, \cg{demonstrating that \SysName can provide significant performance benefits even in highly demanding point cloud applications.} \camera{The one-time coordinate packing and sorting costs on the initial input coordinates account only for 1.3\% and 1.6\% of the total inference time on average for the 32-bit and 64-bit \SysName versions, respectively, as we use RadixSort ~\cite{adinets2022onesweepfastersignificantdigit} for sorting, which has linear 
complexity. In larger scenes, this percentage drops to less than 0.3\%.} Across different datasets and networks, \SysName selects different dataflows with minimal post-processing overheads. \da{In Appendix §\ref{sec:more_gpus}, we evaluate four additional GPUs, and show that \SysName maintains high performance across all GPU architectures, having \SpiraEndToEndAvgAll average speedup.} 

\autoref{tab:resn-thresh}, \autoref{tab:resnl-thresh}, and \autoref{tab:unet-thresh} \camera{report the number of layers for which \SysName{}’s tuning selects different threshold $t$ configurations for the feature computation step in all datasets and networks.  We find that in ScanNet dataset, whose scenes are $\sim$10$\times$ denser on average than those of the other two datasets, more layers with higher threshold values are selected across all networks, which translates to more weight offsets are classified as dense across the layers of the networks.} 

\begin{table}[!htb]
\centering
\vspace{3pt}
\resizebox{0.999\linewidth}{!}{
\begin{tabular}{|l|c|c|c|}
\hline
\rowcolor{teal!16} \camera{\textbf{Number of Layers}} & \camera{\textbf{KITTI}} & \camera{\textbf{ScanNet}} & \camera{\textbf{Waymo}} \\
\hline
\camera{\textbf{Weight-Stationary (t=0)}} & \camera{5} & \camera{0} & \camera{4} \\
\hline
\camera{\textbf{Hybrid-Stationary (t=1)}} & \camera{0} & \camera{0} & \camera{0} \\
\hline
\camera{\textbf{Hybrid-Stationary (t=2)}} & \camera{0} & \camera{4} & \camera{0} \\
\hline
\camera{\textbf{Hybrid-Stationary (t=3)}} & \camera{0} & \camera{0} & \camera{0} \\
\hline
\camera{\textbf{Output-Stationary (t=L1NormMax+1)}} & \camera{16} & \camera{17} & \camera{17} \\
\hline
\end{tabular}
}
\vspace{-10pt}
\caption{\camera{Number of layers selected per threshold configuration for the ResN network using the three evaluated datasets.}}
\label{tab:resn-thresh}
\vspace{6pt}
\end{table}

\begin{table}[!htb]
\centering
\resizebox{0.999\linewidth}{!}{
\begin{tabular}{|l|c|c|c|}
\hline
\rowcolor{teal!16} \camera{\textbf{Number of Layers}} & \camera{\textbf{KITTI}} & \camera{\textbf{ScanNet}} & \camera{\textbf{Waymo}} \\
\hline
\camera{\textbf{Weight-Stationary (t=0)}} & \camera{5} & \camera{1} & \camera{1} \\
\hline
\camera{\textbf{Hybrid-Stationary (t=1)}} & \camera{4} & \camera{0} & \camera{0} \\
\hline
\camera{\textbf{Hybrid-Stationary (t=2)}} & \camera{0} & \camera{4} & \camera{8} \\
\hline
\camera{\textbf{Hybrid-Stationary (t=3)}} & \camera{9} & \camera{9} & \camera{9} \\
\hline
\camera{\textbf{Output-Stationary (t=L1NormMax+1)}} & \camera{2} & \camera{6} & \camera{2} \\
\hline
\end{tabular}
}
\vspace{-10pt}
\caption{\camera{Number of layers selected per threshold configuration for the ResNL network using the three evaluated datasets.}}
\label{tab:resnl-thresh}
\vspace{6pt}
\end{table}

\begin{table}[!htb]
\centering
\resizebox{0.999\linewidth}{!}{
\begin{tabular}{|l|c|c|c|}
\hline
\rowcolor{teal!16} \camera{\textbf{Number of Layers}} & \camera{\textbf{KITTI}} & \camera{\textbf{ScanNet}} & \camera{\textbf{Waymo}} \\
\hline
\camera{\textbf{Weight-Stationary (t=0)}}& \camera{28} & \camera{22} & \camera{25} \\
\hline
\camera{\textbf{Hybrid-Stationary (t=1)}} & \camera{0} & \camera{0} & \camera{0} \\
\hline
\camera{\textbf{Hybrid-Stationary (t=2)}} & \camera{0} & \camera{0} & \camera{0} \\
\hline
\camera{\textbf{Hybrid-Stationary (t=3)}} & \camera{0} & \camera{0} & \camera{0} \\
\hline
\camera{\textbf{Output-Stationary (t=L1NormMax+1)}} & \camera{14} & \camera{20} & \camera{17} \\
\hline
\end{tabular}
}
\vspace{-10pt}
\caption{\camera{Number of layers selected per threshold configuration for the UNet network using the three evaluated datasets.}}
\label{tab:unet-thresh}
\vspace{2pt}
\end{table}

Overall, \SysName \camera{significantly outperforms prior} state-of-the-art engines, demonstrating robust efficiency across diverse point cloud networks, datasets, and GPU architectures.

\vspace{-2pt}
\subsection{Layerwise Performance}
\begin{figure}[h!]  
    \centering 
    \includegraphics[width=\linewidth]{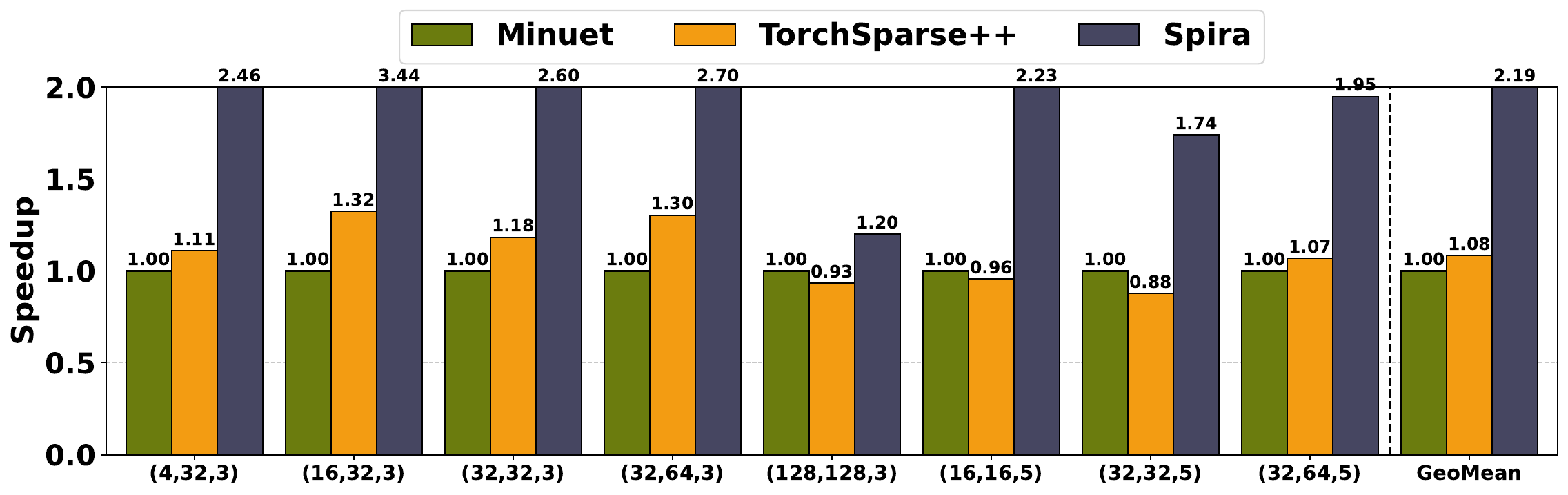}
    \vspace{-25pt}
    \caption{Layerwise speedup of all \SpConv  engines averaged across all datasets, for various \SpConv layers of  $(C_{in}, C_{out}, K)$.}
    \label{fig:layerwise}
\end{figure}

\vspace{-3pt}
Since 64-bit and 32-bit packing perform comparably,
our subsequent evaluations enable \SysName with 32-bit packing. We assess various layers configurations  (input channels $C_{in}$ output channels $C_{out}$ kernel size $K$). For fair comparison with prior work,  we do measure the voxel indexing runtime in \SysName for each evaluated layer. 
\autoref{fig:layerwise} shows layerwise performance of all engines for various layer configurations commonly found in state-of-the-art point cloud 
networks. For each  layer, we report the geometric mean across all datasets. TorchSparse++ and \SysName select the best-performing dataflow.

\begin{figure}[h]  
    \centering 
    \includegraphics[width=\linewidth]{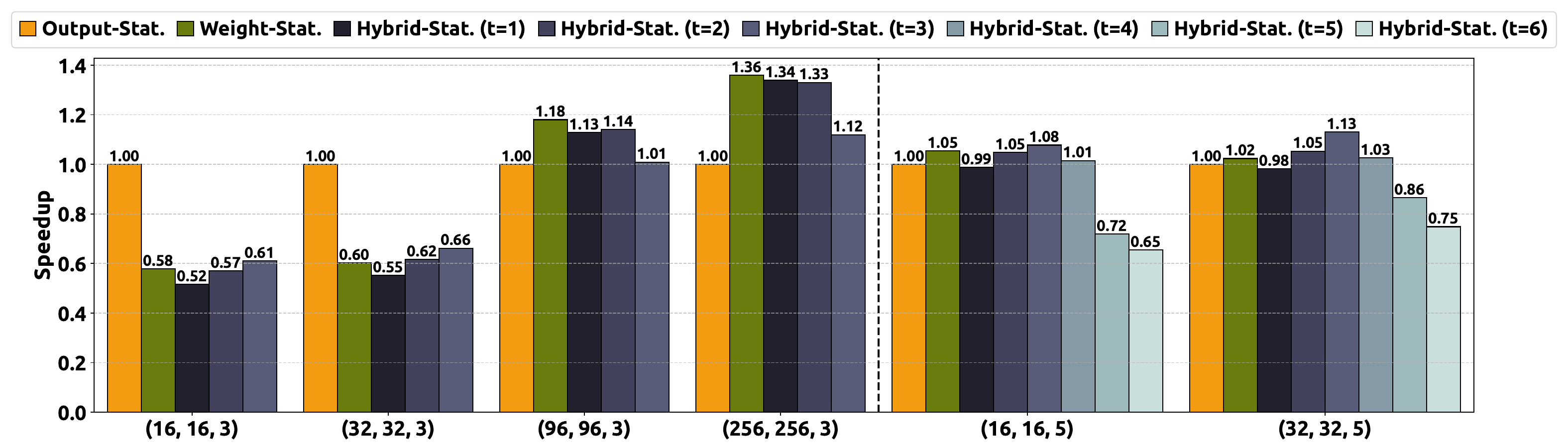}
    \vspace{-24pt}
    \caption{Layerwise speedup of \SysName with various \camera{threshold configurations for the dataflow of the feature computation step.} }
    \label{fig:spira-layerwise}
\end{figure}

We draw two findings. First, TorchSparse++ outperforms Minuet in layers with relatively small channels and kernel sizes (e.g., (16, 32, 3)) because, as we also explain in \autoref{fig:spira-layerwise}, output-stationary dataflow—which Minuet does not support—performs significantly better in such layers. However, in layers with larger channels and kernel sizes, TorchSparse++ achieves comparable performance to Minuet and even surpasses it in some cases.
Second, \SysName consistently outperforms Minuet and TorchSparse++ across all layers, delivering average speedup of \camera{2.19}$\times$ and \camera{2.03}$\times$ (up to \camera{\SpiraLayerMaxThirty} and \camera{2.60}$\times$), respectively. \SysName uses t=3 in hybrid dual-dataflow  for layers  (16, 16, 5) and (32, 32, 5): although hybrid dual-dataflow still incurs post-processing costs, post-processing time is $5.41\times$ and $2.51\times$ lower on average than TorchSparse++ weight-stationary and Minuet, respectively. 


\autoref{fig:spira-layerwise} presents the \SysName layerwise performance using output-, weight-stationary or hybrid dual-dataflow varying the threshold $t$ in submanifold layer configurations with $s_p$=1, \camera{where each triplet of a layer configuration shows the input channels, output channels and kernel size}. 
For $K$=3 and $K$=5, $t$ can get 3 and 6 values, respectively.


We draw two findings. 
First, output-stationary dataflow dominates in layers with smaller channels and kernel sizes. In such layers, weight-stationary incurs high post-processing overhead 
relative to its feature computation time (actual convolution computation is relatively small). Conversely, weight-stationary surpasses output-stationary in layers with larger channels and kernel sizes. For example, in layers with  $K$=5, many weight offsets in kernel map are highly sparse, causing output-stationary to perform numerous unnecessary zero-valued multiplications.
Second, hybrid dual-dataflow performs best in large layers with $K$=5, with speedups up to 1.13$\times$ and 1.11$\times$ over full output-stationary and full weight-stationary, respectively, by combining the strengths of both dataflows. In layers with  $K$=5 (125 weight offsets),
thresholds $t\geq 5$ classify many weight offsets as dense (at least 93), causing high transposition overheads for the dense submatrix.
Instead, threshold $t$=3 provides the best performance: it selects 25 weight offsets as dense (average density 32\%) for output-stationary and 100 as sparse (average density 10\%) for weight-stationary. This split effectively balances compute load across both dataflow executions, justifying the overheads of two separate GPU kernel launches.

\subsection{Performance of Mapping in Voxel Indexing}

\begin{figure}[h!]  
    \centering 
    \includegraphics[width=\linewidth]{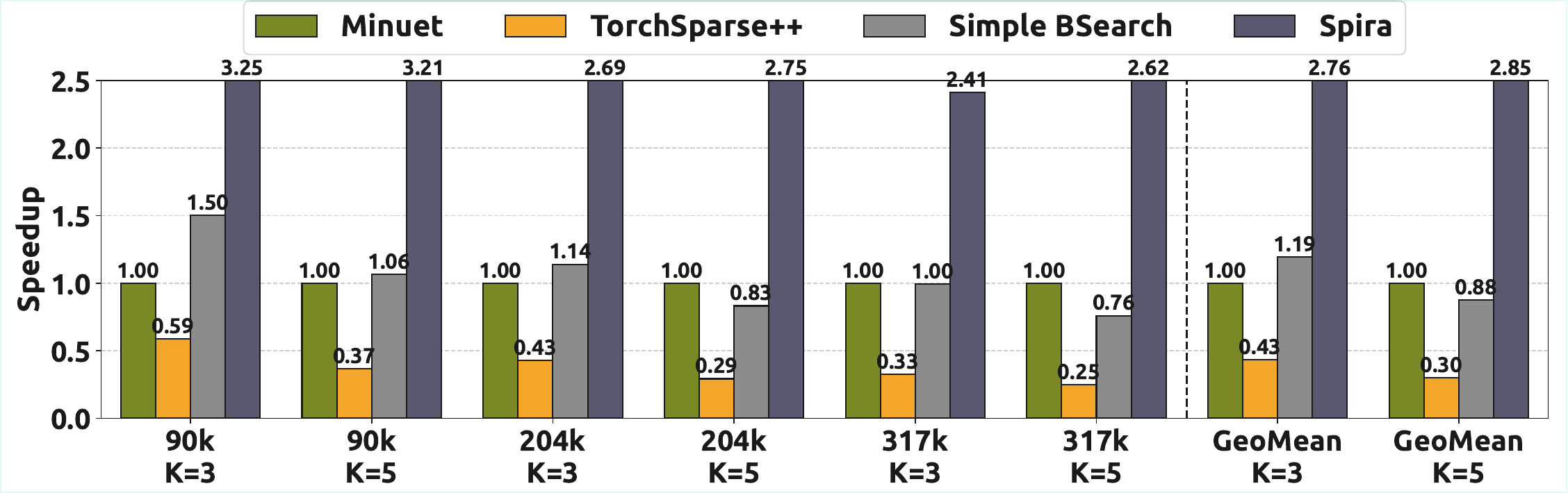}
    \vspace{-26pt}
    \caption{Mapping performance in voxel indexing for all engines across varying input coordinate counts and layer kernel sizes.}
    \label{fig:mapstep}
\end{figure}

\autoref{fig:mapstep} presents mapping performance (pre-processing and search phases of voxel indexing) for all \SpConv engines using real scenes from the evaluated datasets with varying input coordinate counts and layer kernel sizes. We add an additional baseline: lookup operations for kernel map construction are performed via simple binary search (\textbf{Simple BSearch}), which requires no pre-processing. Simple BSearch also includes the 32-bit coordinate packing, as described in §\ref{sec:mechanism-packing}.

We make three key observations. First, Minuet clearly outperforms TorchSparse++ in all cases by 2.83$\times$ on average, as TorchSparse++ causes irregular memory accesses during lookups, while Minuet effectively exploits GPU on-chip caches. 
Second, Simple BSearch outperforms Minuet in smaller scenes (e.g., 90k voxels), but  it scales poorly as the coordinate count and kernel size increase due to poor cache utilization. 
Third, \SysName's z-delta search delivers outstanding performance over all comparison points across all scenes, with speedup up to 2.85$\times$, 9.49$\times$ and 3.45$\times$ over Minuet,  TorchSparse++ and Simple BSearch, respectively. 
Minuet  and TorchSparse++ exhibit notable pre-processing overhead of 8–36.5\% and 18.3–55.5\% of total mapping time, respectively.
Instead, \SysName \emph{completely} eliminates pre-processing and significantly outperforms even packed-native Simple BSearch, which also has \emph{no} pre-processing. \SysName's z-delta search algorithm exploits the integer-valued property to perform intelligent localized searches: for many queries, it requires only a \emph{single comparison} and a \emph{single cache-friendly memory access}, while competing approaches perform multiple comparisons per query and may incur irregular memory accesses. This significantly reduces computational cost and memory access latency, enabling robust performance even as coordinate counts and kernel sizes increase.

\subsection{Performance Breakdown of \SysName's Key Ideas}

\autoref{fig:spira-breakdown-end} \camera{shows the end-to-end inference performance benefits by incrementally enabling \SysName{}'s four key ideas. We present the detailed breakdown on the UNet network. 

\begin{figure}[h!]  
    \centering 
    \includegraphics[width=\linewidth]{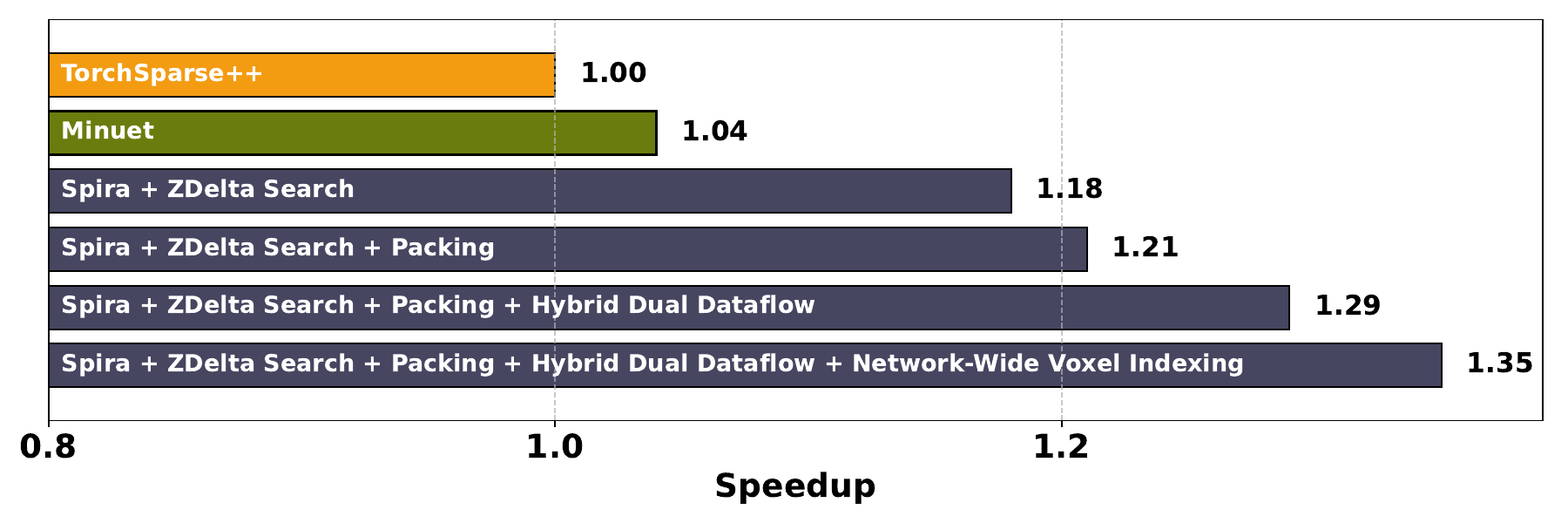}
    \vspace{-28pt}
    \caption{\camera{End-to-end inference performance breakdown of the key ideas of \SysName.}}
    \label{fig:spira-breakdown-end}

\end{figure}

We find that all \SysName's ideas contribute to overall performance, with the largest gains coming from z-delta-search and hybrid dual-dataflow. Specifically, the z-delta-search enables \SysName to achieve a 1.18$\times$ speedup over TorchSparse++, and the hybrid dual-dataflow scheme delivers an additional 1.07$\times$ performance improvement. \SysName's coordinate packing contributes to a $1.02\times$ ($1.06\times$ on average across all networks) speedup. \SysName's network-wide streamed voxel indexing execution further improves end-to-end inference by $1.05\times$ and up to $1.12\times$ across all networks over sequential execution followed by prior works.}

\begin{figure}[h]  
    \centering 
    \includegraphics[width=\linewidth]{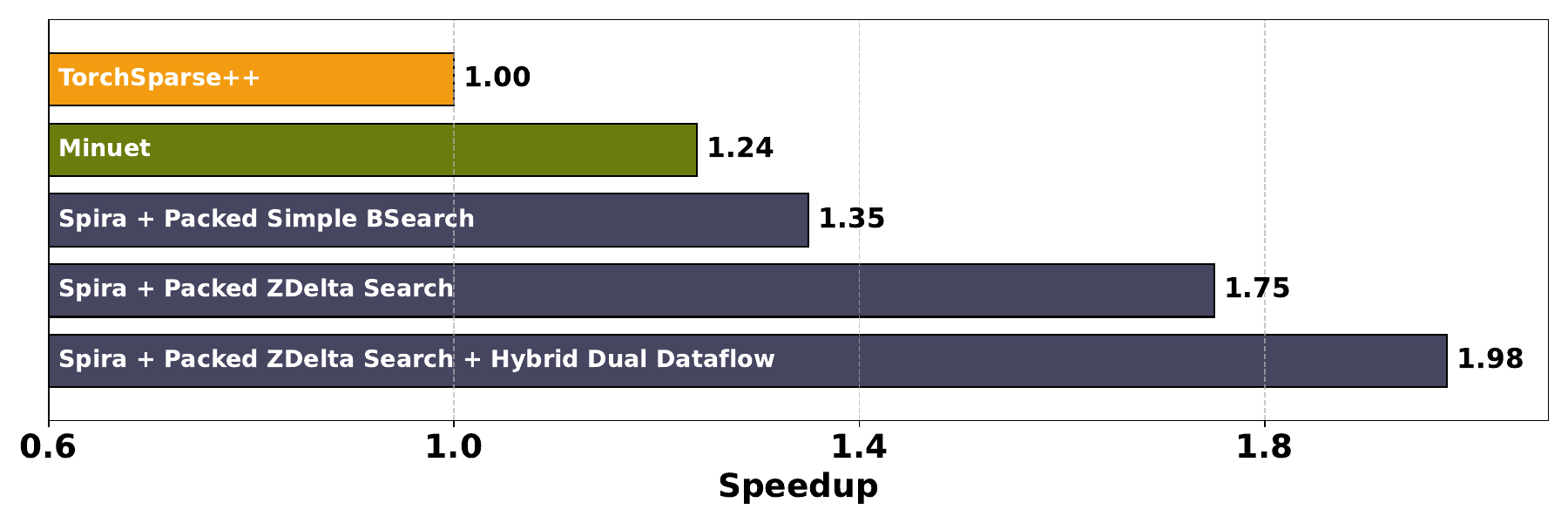}
    \vspace{-28pt}
    \caption{Layerwise performance breakdown of \SysName key ideas.}
    \label{fig:spira-breakdown-layerwise}

\end{figure}

\autoref{fig:spira-breakdown-layerwise} \camera{shows the layerwise performance improvements by incrementally adding \SysName's optimizations}: (1) 32-bit packed-native voxel indexing on Simple BSearch with output-stationary dataflow, (2) replacing Simple BSearch with \SysName's z-delta search, and (3) replacing output-stationary with \SysName's adaptive hybrid dual-dataflow. 
We evaluate a common layer of $(C_{in}, C_{out}, K)$=(32,32,5), which appears in 4 of 20 layers in ResNL, and include the voxel indexing runtime. 

The most substantial speedup comes from \camera{replacing Simple BSearch with our z-delta search algorithm (resulting $1.75\times$ speedup over TorchSparse++)}, demonstrating the effectiveness of our \emph{one-shot} design (no pre-processing) and highly localized search scheme that is both compute- and memory-efficient.

\begin{figure}[h]
    \centering
    \includegraphics[width=0.9\linewidth]{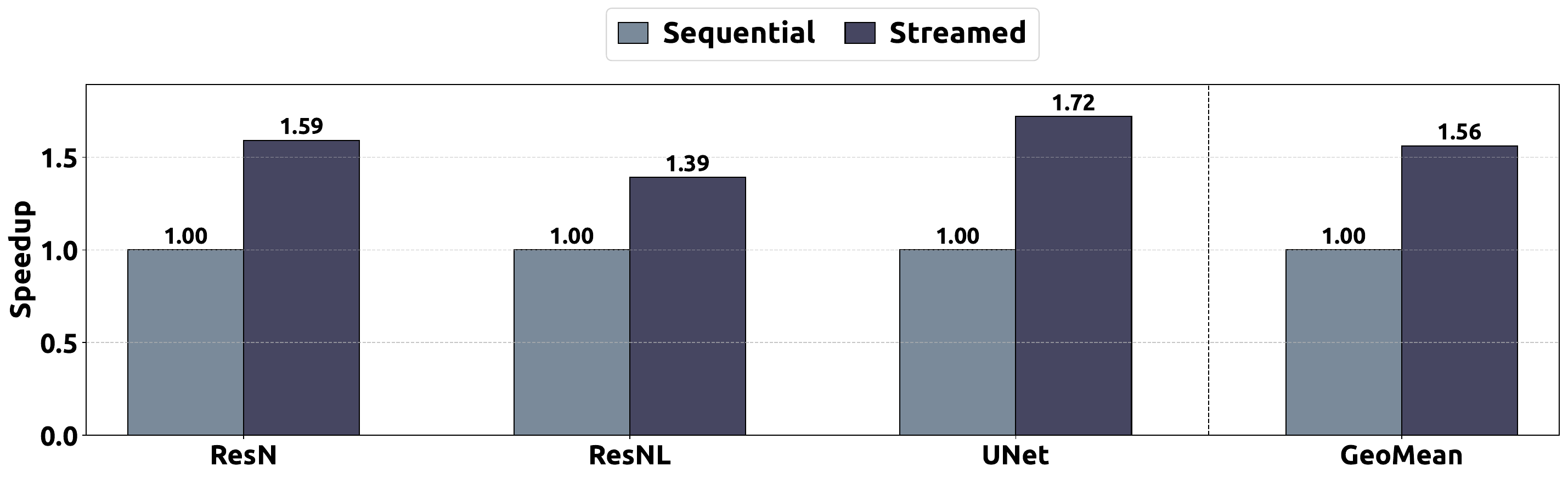}
    \vspace{-15pt}  
    \caption{\SysName with sequential \camera{versus} streamed voxel indexing execution.
    }
    \label{fig:spira-voxel-indexing}
\end{figure}

\autoref{fig:spira-voxel-indexing} shows the performance benefits of \SysName's network-wide voxel indexing optimization. We measure total voxel indexing time to create kernel maps for all layers of each of the three networks, and evaluate either sequential execution (i.e., the approach followed by prior works~\cite{Choy2019Minkownski,Tang2023TorchSparse++,Yang2024Minuet}) or \SysName's streamed (concurrent) execution. 
\SysName’s network-wide streamed execution improves total voxel indexing latency by up to $1.72\times$ over sequential execution.

\subsection{Ablation Studies}

\begin{figure}[h!]  
    \centering 
    \includegraphics[width=\linewidth]{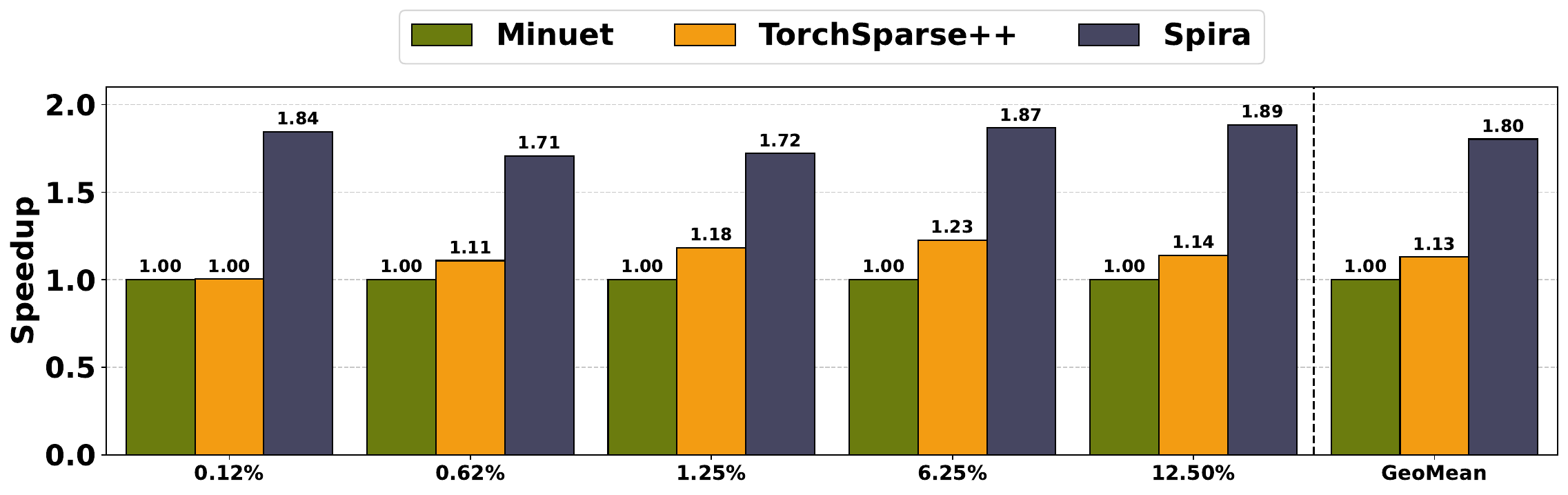}
    \vspace{-25pt}
    \caption{\da{End-to-end inference performance of all \SpConv engines averaged across all networks for scenes of different densities that may not follow the neighboring property existing in real datasets.}}
    \label{fig:sparsity_ablation}
\end{figure}

\camera{We generate synthetic scenes with randomly distributed \da{non-zero} voxels inside a [200, 200, 200] volume to evaluate the \SpConv engines on voxel data that do not necessarily follow the neighboring property of real-world point clouds. \autoref{fig:sparsity_ablation} shows the average end-to-end inference performance of all engines across all networks \da{on synthetic scenes whose density ranges from 0.12\% to 12.50\%}. \da{Even on synthetic scenes that may not follow the neighboring property of real-world point clouds}, \SysName outperforms Minuet and TorchSparse++ by on average 1.80$\times$ and 1.59$\times$, respectively.}

\camera{To evaluate how the \SpConv engines scale \da{when increasing the input coordinate count}, we generate synthetic scenes with a fixed density of $\sim$1.25\%, by increasing both the number of non-zero voxels and the 3D space volume. Specifically,  we select a volume of size $[n, n, 200]$, and increase the value $n$ of the dimensions $x$ and $y$, then we randomly generate  non-zero voxels inside, such that the scene matches the target density. \autoref{fig:large scale} shows the average end-to-end inference performance of all engines on ResN network, when varying the number of \da{non-zero} voxels from $10^4$ to $5 \times 10^6$ while keeping scene density fixed (scene volume increases). 

\begin{figure}[h!]  
    \centering 
    \includegraphics[width=\linewidth]{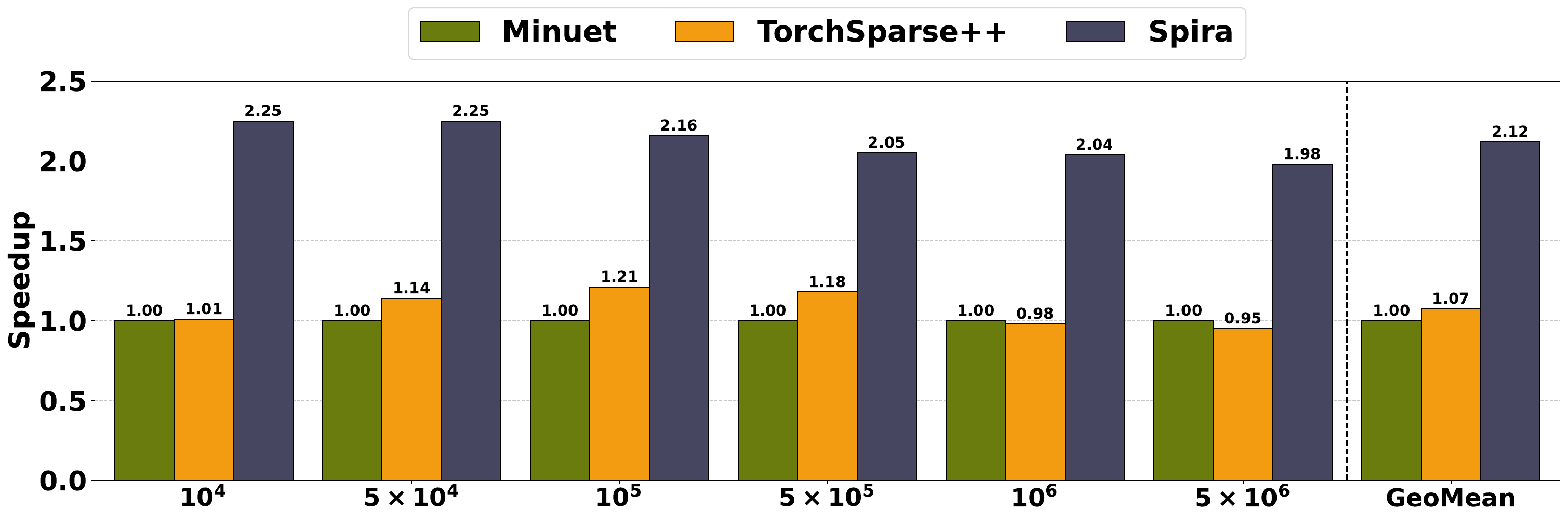}
    \vspace{-25pt}
    \caption{\camera{End-to-end inference performance of all \SpConv engines on ResN network for synthetic scenes of fixed density and different number of non-zero voxels.}}
    \label{fig:large scale}
\end{figure}

\SysName effectively scales on larger scenes with large number of non-zero voxels $(\geq 10^6)$, outperforming existing \SpConv engines by on average 2.05$\times$ (up to 2.25$\times$).}

\camera{In Appendix §\ref{sec:mem_table}, we also evaluate  the peak memory footprint and performance of all \SpConv engines during end-to-end inference, while varying the kernel size and batch size. Our results demonstrate that \SysName has similar memory footprint to prior engines, while achieving large performance gains even for larger batch sizes and kernel sizes.}



%% file: content/related_work.tex
\vspace{-4pt}
\section{Other Related Work}
\vspace{-4pt}
\noindent\textbf{Point Cloud Accelerators.} Prior works~\cite{Feng2020Mesorasi,Han2024BitNN,Zhang2021PointX,Feng2022Crescent,Lin2021PointAcc,lin2024voxel,lyu2023spocta} propose application-specific accelerators for point cloud networks. However, only a few accelerators~\cite{Lin2021PointAcc,lin2024voxel,lyu2023spocta} support the 3D \SpConv. These works have custom microarchitecture designs and rely on simulators for  evaluation, which limits their immediate practical deployment. Instead, \SysName{} is a software runtime engine  that directly runs on commodity high-end and edge GPUs, provides comprehensive evaluations on real systems, and  enables immediate deployment in real \SpConv applications.

\noindent\textbf{Deep Learning Compilers.} Deep Learning (DL) compilers optimize dense tensor algebra~\cite{feng2023tensorir,Chen2018TVM,Ding2023Hidet,xing2022bolt} and sparse tensor algebra kernels~\cite{Gupta2025SPLAT,Liu2025CROSS,Ahrens2025Finch,Du2025SRSparse,won2023unified,Ye2023SparseTIR,Kjolstad2017TACO}. Although some  sparse DL compilers could be used to optimize \SpConv, these compilers do not integrate the optimizations proposed in \SysName. \SysName's ideas can work synergistically with existing sparse DL compilers to significantly improve performance  in \SpConv executions.

%% file: content/conclusion.tex
\vspace{-4pt}
\section{Conclusion}
\vspace{-4pt}

We introduce \SysName, the first voxel-property-aware \SpConv engine for GPUs. \SysName proposes the one-shot z-delta search algorithm that has low computational cost and high \camera{data} locality, integrates packed-native voxel indexing that significantly reduces data accesses, employs network-wide parallelization in voxel indexing kernels of all network's \SpConv layers to increase execution parallelism, and provides flexible hybrid dual-dataflow feature computation that adapts to layer characteristics. \SysName significantly outperforms prior state-of-the-art \SpConv engines by \camera{\SpiraEndToEndAvgAll} on average and up to \camera{\SpiraEndToEndMaxAll} in end-to-end inference across various point cloud networks, real-world datasets, and GPUs. 
\camera{We hope our work encourages further research studies on sparse operators of emerging ML models and structural properties of 3D data.}


%% file: content/acknowledgements.tex
\vspace{-5pt}
\section{Acknowledgements}
\vspace{-5pt}
We \camera{thank the MLSys 2026 reviewers for their valuable feedback and the members of the SPIN research group and the CSLab research group for the stimulating and inclusive research environments they provide.
We thank the IT staff at the Max Planck Institute for Software Systems for technical support.
We thank the Visual Computing and Artificial Intelligence department at the Max Planck Institute for Informatics for lending us the Jetson Orin AGX platform.}

%% file: content/appendix.tex
\clearpage
\section{Additional Evaluation Results}

\subsection{End-To-End Inference on More GPUs}\label{sec:more_gpus}

\begin{figure}[h!]  
   \vspace{6pt}
    \centering 
    \includegraphics[width=\linewidth]{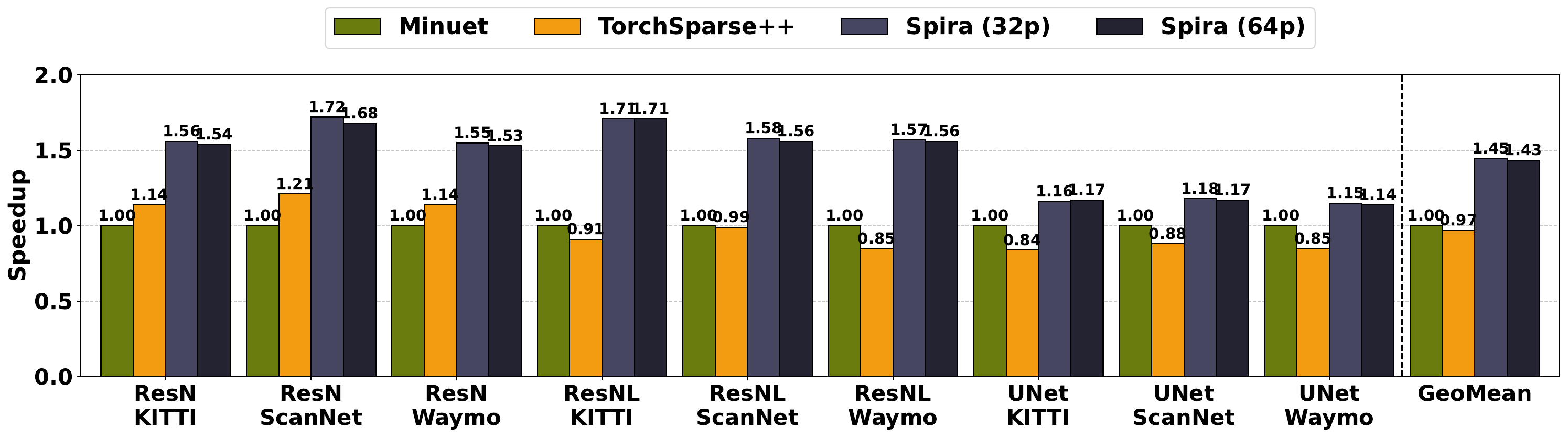}
    \vspace{-23pt}
    \caption{\camera{End-to-end inference performance of all \SpConv engines using various point cloud networks and datasets on GTX 1060 GPU.}}
    \label{fig:end-to-end-1060}
    \vspace{6pt}
\end{figure}

\begin{figure}[h!]  
    \centering 
    \includegraphics[width=\linewidth]{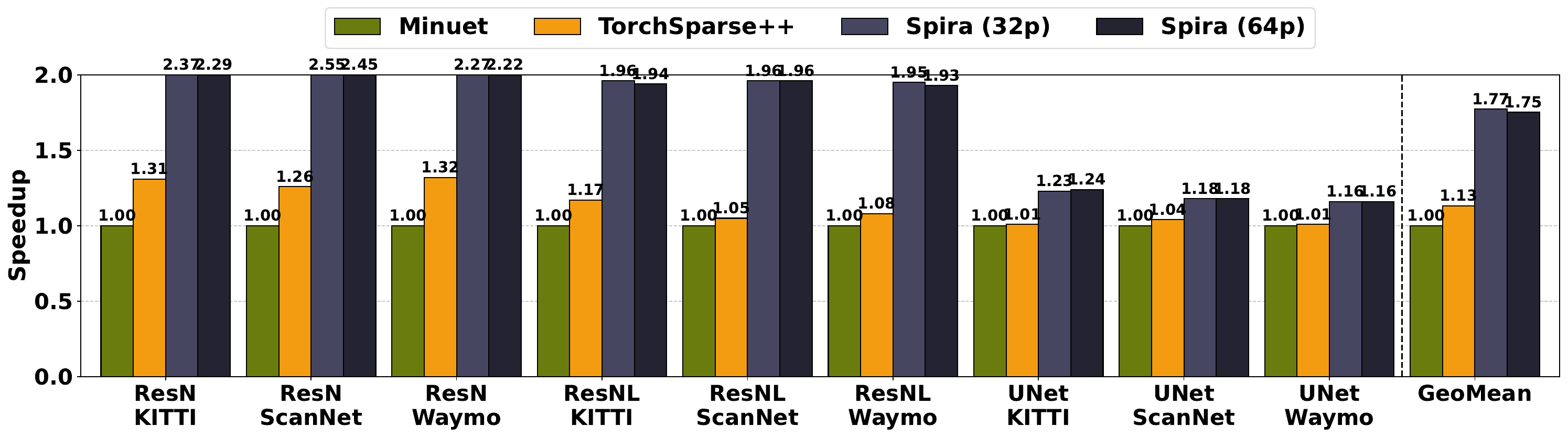}
    \vspace{-23pt}
    \caption{\camera{End-to-end inference performance of all \SpConv engines using various point cloud networks and datasets on Quadro RTX 5000 GPU.}}
    \label{fig:end-to-end-quadro}
     \vspace{6pt}
\end{figure}

\begin{figure}[h!]  
    \centering 
    \includegraphics[width=\linewidth]{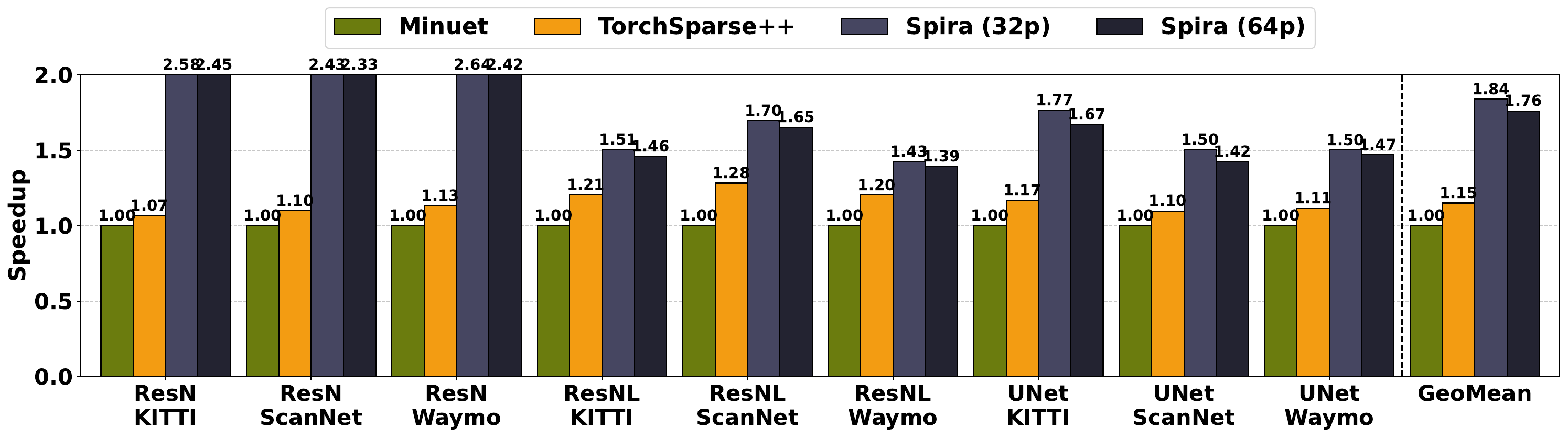}
    \vspace{-23pt}
    \caption{\camera{End-to-end inference performance of all \SpConv engines using various point cloud networks and datasets on H100 GPU.}}
    \label{fig:end-to-end-h100}
     \vspace{8pt}
\end{figure}

\begin{figure}[h!]  
    \centering 
    \includegraphics[width=\linewidth]{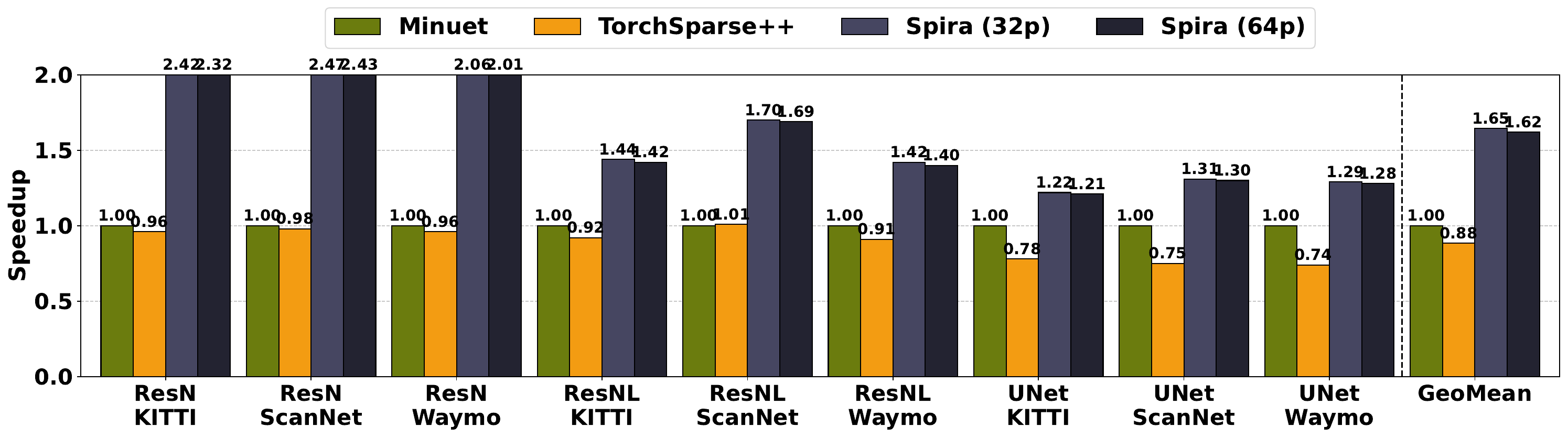}
    \vspace{-23pt}
    \caption{\camera{End-to-end inference performance of all \SpConv engines using various point cloud networks and datasets on Jetson Orin AGX platform.}}
    \label{fig:end-to-end-orin}
     \vspace{6pt}
\end{figure}

\autoref{fig:end-to-end-1060}, \autoref{fig:end-to-end-quadro}, \autoref{fig:end-to-end-h100}, and \autoref{fig:end-to-end-orin} \camera{show the end-to-end inference performance of all \SpConv engines across different networks and datasets, measured on GTX 1060, Quadro RTX 5000, H100, and Jetson Orin AGX GPUs, respectively. For these benchmarks, all evaluations on the GTX 1060 are performed in 32-bit float precision, since 16-bit float precision in this architecture is inefficiently supported, i.e., it achieves 64$\times$ lower compute throughput than that of 32-bit float.}

\camera{\SysName significantly outperforms Minuet and TorchSparse++ across all GPUs, with 32-bit packing delivering consistent average speedups of 1.47$\times$, 1.67$\times$, 1.71$\times$, and 1.75$\times$ on GTX 1060, Quadro RTX 5000, H100, and Jetson Orin AGX, respectively.
These results demonstrate \SysName{}’s ability to provide robust performance gains regardless of the underlying GPU architecture.}

 \vspace{-18pt}
\camera{\subsection{Memory Footprint Evaluation}\label{sec:mem_table}

\autoref{tab:compact_memory_speedup} presents the peak memory footprint (including the memory footprint of kernel maps, weights, and features) and performance of all \SpConv engines measured during end-to-end inference on the Waymo Dataset. We evaluate UNet (42 layers) with large kernel size (\textbf{K}) up to 13 and batch size (\textbf{B}) up to 4 using the RTX 3090 GPU (24GB).

\begin{table}[!htb]
\centering
 \vspace{6pt}
\resizebox{\linewidth}{!}{
\vspace{2pt}
\begin{tabular}{|l|c|c|c|}
\hline
\rowcolor{teal!16}\textbf{\camera{Configuration}} & \textbf{\camera{TorchSparse++}} & \textbf{\camera{Minuet}} & \textbf{\camera{\SysName}} \\
\hline
\textbf{\camera{K=3; \hspace{8pt} B=1}}  & \camera{353MB / 1.00} & \camera{405MB / 1.06$\times$} & \camera{360MB / 1.33$\times$} \\
\hline
\textbf{\camera{K=7; \hspace{8pt} B=1}}  & \camera{1.6GB / 1.00} & \camera{1.7GB / 1.30$\times$} & \camera{1.4GB / 2.14$\times$} \\
\hline
\textbf{\camera{K=13; \hspace{4pt} B=1}} & \camera{9.2GB / 1.00} & \camera{9.6GB / 1.15$\times$} & \camera{9GB / 2.20$\times$} \\
\hline
\hline
\textbf{\camera{K=3; \hspace{8pt} B=4}} & \camera{1.1GB / 1.00} & \camera{1.3GB / 1.05$\times$} & \camera{1.1GB / 1.28$\times$} \\
\hline
\textbf{\camera{K=7; \hspace{8pt} B=4}}  & \camera{3GB / 1.00} & \camera{3.5GB / 1.28$\times$} & \camera{2.7GB / 1.75$\times$} \\
\hline
\textbf{\camera{K=13; \hspace{4pt} B=4}} & \camera{17GB / 1.00} & \camera{OOM / --} & \camera{16GB / 1.81$\times$} \\
\hline
\end{tabular}
}
\vspace{-8pt}
\caption{\camera{Peak memory footprint and performance of all \SpConv engines in end-to-end inference for different kernel size (\textbf{K}) and batch size (\textbf{B}) configurations.}}
\label{tab:compact_memory_speedup}
 \vspace{2pt}
\end{table}

Our results show that the peak memory footprint of all \SpConv engines is similar. \da{While \SysName pre-computes all kernel maps upfront at the network start, TorchSparse++ and Minuet generate them progressively, as layers are executed, and retain them in memory to reuse them across \SpConv layers that share the same kernel maps (See §\ref{sec:mechanism-network-voxel-indexing}). Hence, all \SpConv engines converge to have similar memory footprints for kernel maps, as well as all other inference data including features and weights, which remain identical across the engines. Additionally, \SysName outperforms Minuet and TorchSparse++ by up to 1.30$\times$ and 2.20$\times$, respectively.} 
In these experiments, the minimum and maximum kernel map memory footprint of \SysName is 19MB (K=3 \& B=1) and 6.4GB (K=13 \& B=4), respectively.}

\section{Artifact Appendix}

\subsection{Abstract}

The Artifact Appendix describes how to reproduce the main results of this paper. It includes the source code of \SysName, benchmark scripts, and step-by-step instructions for the key evaluation results.  The experiments require an NVIDIA GPU with an up-to-date NVIDIA driver installed. 
We provide a \texttt{README.md} file that describes the required hardware and software dependencies and provides step-by-step instructions. 
Note that the datasets used in our evaluations are licensed. This artifact is used to support our major claims (See §\ref{sec:claims}), demonstrating \SysName's performance benefits in \autoref{fig:end-to-end-3090}, \autoref{fig:end-to-end-A100}, \autoref{fig:layerwise}, \autoref{fig:mapstep}, and \autoref{fig:sparsity_ablation}.
We expect the full evaluation pipeline, including setup to take approximately 2–3 hours.
We also provide a Dockerfile to automatically set up the runtime environment for running the artifact. We \textbf{recommend} using the docker environment for running the experiments.

\subsection{Artifact check-list (meta-information)}

\begin{itemize}[topsep=0pt,leftmargin=12pt,nosep,partopsep=0pt]
\item {\bf Program: } \textit{Spira\_Artifact}: In this artifact, we compile and evaluate three comparison points Minuet, TorchSparse++ and \SysName.
\item {\bf Compilation: } CMake build system; GNU compilers (gcc/g++); NVIDIA CUDA compiler (nvcc); PyTorch. The artifact is compiled as a Docker image for ease of deployment and reproducibility.
\item {\bf Data set:} SemanticKITTI (KITTI), ScanNet and Waymo. All datasets are licensed. 
\item {\bf Run-time environment:} Linux Ubuntu 22.04 with Python 3.10, requiring CUDA 12.4.1 and PyTorch 2.5.0, all provided via a Docker image.
\item {\bf Hardware:} A system with an NVIDIA GPU device with a minimum compute capability of 7.5 and at least 16GB GPU memory should be used to validate the results.
\item {\bf Metrics:} Execution time in milliseconds normalized as relative performance speedup.
\item {\bf Output:} Output files containing raw results. Figures similar to \autoref{fig:end-to-end-3090}, \autoref{fig:end-to-end-A100}, \autoref{fig:layerwise}, \autoref{fig:mapstep}, and \autoref{fig:sparsity_ablation} of the main paper.
\item {\bf Experiments:} End-to-end inference performance, layerwise performance, mapping performance, scene density ablation study, correctness check.
\item {\bf How much disk space required (approximately)?: }256 GB. 
\item {\bf How much time is needed to prepare workflow (approximately)?: }60 minutes (build the code and download datasets). 
\item {\bf How much time is needed to complete experiments (approximately)?: }90 minutes. 
\item {\bf Publicly available?: }Yes.
\item {\bf Code licenses (if publicly available)?: } Apache 2.0.
\item {\bf Data licenses (if publicly available)?: } ScanNet: ScanNet Terms of Use; Waymo Open Dataset: Waymo Open Dataset License; SemanticKITTI: CC BY-NC-SA 4.0.
\item {\bf Archived (provide DOI)?: }10.5281/zenodo.18879475. 
\end{itemize}

\subsection{Description}

\subsubsection{How to Access} \label{sec:appendix-source-code}
Download the compressed file Spira\_Artifact.zip from the Zenodo archive \href{https://doi.org/10.5281/zenodo.18879475}{https://doi.org/10.5281/zenodo.18879475}  or our GitHub repository at \href{https://github.com/SPIN-Research-Group/Spira}{https://github.com/SPIN-Research-Group/Spira}.

\subsubsection{Hardware Dependencies}

The artifact should be tested on a host machine with:

\begin{itemize}[topsep=0pt, leftmargin=12pt, partopsep=0pt, parsep=0pt, itemsep=0pt, noitemsep, before=\vspace{-\parskip}]
\item x86-64 CPU with at least 64GB main memory and 256GB disk storage.
\item NVIDIA GPU device with a compute capability (SM) of 7.5–9.0 and at least 16GB GPU memory.
\end{itemize}


\subsubsection{Software Dependencies}

The artifact requires the following software for installation:

\begin{itemize}[topsep=0pt, leftmargin=12pt, partopsep=0pt, parsep=0pt, itemsep=0pt, noitemsep, before=\vspace{-\parskip}]
\item Ubuntu 22.04 (or newer)
\item Python 3.10
\item GNU compilers (gcc/g++) 11.4.0 (\textbf{strict} requirement)
\item CUDA 12.4.1 (or newer)
\item PyTorch 2.5.0 (or newer)
\item CMake 3.27.0 (\textbf{strict} requirement)
\item Pybind 2.11 (\textbf{strict} requirement)
\item libsparsehash-dev 2.0.3 (or newer)
\item libopenblas-dev  0.3.20 (or newer)
\end{itemize}

To simplify setup and ensure reproducibility, we \textbf{recommend} building the artifact as a Docker image. The Dockerfile installs all necessary software dependencies, including additional Python packages for dataset preparation and figure generation. To build and run the artifact as a Docker image, users should have a Linux-based operating system with an up-to-date NVIDIA driver (supporting at least CUDA 12.4), Docker Engine and NVIDIA Container Toolkit installed. For reference, our environment uses: 
\begin{itemize}[topsep=0pt, leftmargin=12pt, partopsep=0pt, parsep=0pt, itemsep=0pt, noitemsep, before=\vspace{-\parskip}]
\item Debian GNU/Linux 12
\item NVIDIA driver 550.54.14
\item Docker Engine 29.2
\item NVIDIA Container Toolkit 1.13.5
\end{itemize}

\subsubsection{Data Sets}

For our evaluation, we use scenes from 3 licensed real-world datasets. We provide detailed instructions for getting access, downloading and preparing the datasets in the \texttt{README.md} file of the artifact. The real-world datasets are the following:

\begin{itemize}[topsep=0pt, leftmargin=12pt, partopsep=0pt, parsep=0pt, itemsep=0pt, noitemsep, before=\vspace{-\parskip}]
\item SemanticKITTI (KITTI)
\item ScanNet
\item Waymo
\end{itemize}

\subsection{Installation}

Download the zip file containing the artifact source code in §\ref{sec:appendix-source-code}.
We provide detailed instructions in the \texttt{README.md} file under the root of source code directory to build and install \SysName and baselines.

We provide a Dockerfile to setup the runtime environment for all
the experiments.
\begin{enumerate}[topsep=0pt,leftmargin=12pt,nosep,partopsep=0pt]
    \item Install Docker Engine following the instructions provided in
\href{https://docs.docker.com/engine/install/ubuntu/}{https://docs.docker.com/engine/install/ubuntu/}.
\item Install NVIDIA Container Toolkit following the instructions provided in
\href{https://docs.nvidia.com/datacenter/cloud-native/container-toolkit/latest/install-guide.html}{https://docs.nvidia.com/datacenter/cloud-native/container-toolkit/latest/install-guide.html}.
    \item Download the source code of Spira\_Artifact.
    \item Export the variable \texttt{CUDA\_ARCHS} with the targeted GPU architecture for evaluation and build the Docker image by executing the following command at the root directory of the source code: 
    \begin{tcolorbox}[
  colback=gray!10, 
  colframe=gray!50, 
  boxrule=0.4pt, 
  left=4pt, right=4pt, top=2pt, bottom=2pt,
  halign=flush left,
  sharp corners
]
\ttfamily
\$ docker build --build-arg CUDA\_ARCHS=\$CUDA\_ARCHS -t spira .
\end{tcolorbox}
\end{enumerate}

\subsection{Experiment workflow}

The artifact contains 5 experiments to evaluate the performance and correctness of the \SpConv engines, with each experiment executed by a script located in the \texttt{automate/} directory. Each script runs the experiment, parses the raw results, and generates a corresponding figure. The first four experiments 1-4 measure execution time and save the raw results under the \texttt{results/} directory, and generate figures under the \texttt{figures/} directory. Next, we describe in detail how to run each experiment.

\noindent\textit{1. End-to-End Inference Performance: } The following script evaluates the end-to-end inference performance of all \SpConv engines across different datasets and networks (\autoref{fig:end-to-end-3090} and \autoref{fig:end-to-end-A100}):
\begin{tcolorbox}[colback=gray!10, colframe=gray!50, boxrule=0.4pt, left=4pt, right=4pt, top=2pt, bottom=2pt]
\texttt{\$ bash automate/end\_to\_end.sh}
\end{tcolorbox}

\noindent\textit{2. Layerwise Performance: } The following script evaluates the layerwise performance of all \SpConv engines averaged across all datasets for different \SpConv  layer configurations (\autoref{fig:layerwise}):
\begin{tcolorbox}[colback=gray!10, colframe=gray!50, boxrule=0.4pt, left=4pt, right=4pt, top=2pt, bottom=2pt]
\texttt{\$ bash automate/layerwise.sh}
\end{tcolorbox}

\noindent\textit{3. Mapping Performance: } The following script evaluates the mapping performance in voxel indexing step of all \SpConv engines for various input coordinate counts and layer kernel sizes (\autoref{fig:mapstep}):
\begin{tcolorbox}[colback=gray!10, colframe=gray!50, boxrule=0.4pt, left=4pt, right=4pt, top=2pt, bottom=2pt]
\texttt{\$ bash automate/mapping.sh}
\end{tcolorbox}

\noindent\textit{4. Scene Density Ablation Study: } The following script evaluates end-to-end inference performance averaged across all networks for synthetic scenes of varying sparsity (\autoref{fig:sparsity_ablation}):
\begin{tcolorbox}[colback=gray!10, colframe=gray!50, boxrule=0.4pt, left=4pt, right=4pt, top=2pt, bottom=2pt]
\texttt{\$ bash automate/ablation.sh}
\end{tcolorbox}

\noindent\textit{5. Correctness: } The following command verifies the correctness of all \SpConv engine outputs (coordinates and features), including all threshold selections for \SysName:
\begin{tcolorbox}[colback=gray!10, colframe=gray!50, boxrule=0.4pt, left=4pt, right=4pt, top=2pt, bottom=2pt]
\texttt{\$ bash automate/correctness.sh}
\end{tcolorbox}

\noindent\textit{All: } The following command will execute \textit{all} experiments and generate \textit{all} figures: 
\begin{tcolorbox}[colback=gray!10, colframe=gray!50, boxrule=0.4pt, left=4pt, right=4pt, top=2pt, bottom=2pt]
\texttt{\$ bash automate/run\_all.sh}
\end{tcolorbox}

\subsection{Evaluation and expected result}\label{sec:claims}

\textbf{Major Claims}. For each of the first four experiments 1-4, we expect the reproduced results to  be similar to those reported in the paper, given the same input configuration. We next clarify our major claims:

\begin{enumerate}[topsep=0pt, leftmargin=12pt, partopsep=0pt, parsep=0pt, itemsep=0pt, noitemsep, before=\vspace{-\parskip}]
\item \SysName achieves significant end-to-end point cloud inference speedup over prior state-of-the-art \SpConv engines on modern GPUs. As indicative results, we report speedups in the range of 1.5$\times$--2.1$\times$ on the RTX 3090 and A100 GPUs. 
\item \SysName achieves significant layerwise speedup over prior state-of-the-art \SpConv engines on modern GPUs. As indicative results, we report speedups in the range of 1.9$\times$--2.6$\times$ on the RTX 3090 and A100 GPUs across diverse \SpConv layer configurations.
\item \SysName's z-delta search delivers significant mapping part acceleration over Minuet, TorchSparse++, and Simple BSearch on modern GPUs. As indicative results, we report speedups in the range of 2.7$\times$--7.8$\times$ on the RTX 3090 and A100 GPUs for the mapping part of voxel indexing step.
\item When evaluated on randomly generated scenes with densities ranging from 0.12\% to 12.50\%, \SysName achieves significant end-to-end point cloud inference speedup over prior state-of-the-art \SpConv engines on modern GPUs. As indicative results, we report speedups in the range of 1.5$\times$--3.1$\times$ on the RTX 3090 and A100 GPUs.
\end{enumerate}

\subsection{Experiment customization}

Each bash script can be configured to modify the input arguments (e.g., dataset, network, scenes) of the corresponding experiment. The input arguments are defined at the beginning of each script and can be modified to test different input configurations. For example, in the end-to-end inference performance script, lines 3--5 can be modified to test different datasets, models (i.e., different networks) and/or \SpConv libraries (i.e., different engines):

\begin{tcolorbox}[colback=gray!10, colframe=gray!50, boxrule=0.4pt, left=4pt, right=4pt, top=2pt, bottom=2pt]
\begin{minipage}[t]{\linewidth}
\texttt{DATASETS=(...)}\\
\texttt{MODELS=(...)}\\
\texttt{LIBS=(...)}
\end{minipage}
\end{tcolorbox}